\documentclass[a4paper,fleqn]{article}
 
\usepackage[numbers]{natbib}
 
\usepackage{amsmath, amssymb, amsthm}   
\usepackage{mathtools}                  
\usepackage{graphicx}                   
\usepackage{booktabs}                   
\usepackage{multirow}                   
\usepackage{hyperref}                   
\usepackage{cleveref}                   
\usepackage{xcolor}                     
\usepackage{algorithm}                  
\usepackage{algpseudocode}              
\usepackage{subcaption}                 
\usepackage{todonotes}                  
\usepackage{microtype}                  
\usepackage{geometry}                   
\usepackage{enumitem}
\usepackage{soul}
 
\usepackage{sidecap}
\sidecaptionvpos{figure}{t}  

\newcommand{\eps}{\varepsilon}           
\newcommand{\dist}{\mathcal{D}}          
\newcommand{\abs}{\mathcal{A}}           
\newcommand{\recov}{\mathcal{R}}         
\newcommand{\degrad}{\mathcal{G}}        
\newcommand{\norm}[1]{\left\|#1\right\|} 
\newcommand{\R}{\mathbb{R}}              
 
\begin{document}
 
 
\title{Towards a Resilience-Theoretic Foundation for Adversarial Robustness in Industrial Control System Anomaly Detection%
\thanks{This work was supported by the European Union's Horizon Research and Innovation Programme under GA No. 101119681 (project ResilMesh).}}
 
\author{
Branka Stojanović\textsuperscript{1,*},~
Andreas Flatscher\textsuperscript{1},~
Michael Somma\textsuperscript{1,2}
\\[6pt]
\small\textsuperscript{1}JOANNEUM RESEARCH Forschungsgesellschaft mbH, Austria\\
\small\textsuperscript{2}TU Graz, Institute for Technical Informatics, Austria
\\[6pt]
\small\textsuperscript{*}Corresponding author: \texttt{branka.stojanovic@joanneum.at}\\
\small\texttt{andreas.flatscher@joanneum.at}, \texttt{michael.somma@joanneum.at}
}
 
\date{}
 
\maketitle
 
 
\begin{abstract}
Anomaly-based intrusion detection systems in industrial control systems (ICS) and operational technology (OT) environments are increasingly required to meet formal resilience criteria: absorbed adversarial disturbances, graceful degradation under sustained attack, and certified system-level guarantees. Existing resilience frameworks for cyber-physical systems define absorb-recover-adapt trajectories at the architectural level but do not treat machine learning anomaly detectors as first-class
components, leaving a gap between component-level robustness evaluation and system-level resilience certification.
 
In this paper, we establish that adversarial robustness in ICS anomaly detection is a
specific instantiation of system resilience, and formalise this connection by mapping four resilience constructs, i.e. disturbance class, absorption capacity, recovery trajectory, and degradation function, onto the adversarial machine learning setting. We derive a compositional resilience bound for heterogeneous ICS detection networks, showing that the binding constraint on system-level resilience is the coupling-adjusted
absorption capacity of each node along the attack path, not the per-node capacity -- so the binding node need not be the weakest one. Empirical validation on the BATADAL water distribution system benchmark demonstrates that the resulting metrics surface operationally significant phenomena invisible to standard benchmarks: the absorption-degradation divergence under adversarial training, and the paradox that hardening the binding node in isolation reduces system-level resilience. Implications for ICS architecture design and certification standards are discussed.
\end{abstract}
 

\bigskip

\section{Introduction}
\label{sec:intro}
 
Industrial control systems (ICS) and operational technology (OT) environments are among the most consequential deployment contexts for machine learning: anomaly detectors operate continuously on sensor streams, protect physical processes from adversarial manipulation, and underpin the security posture of critical infrastructure ranging from water treatment to power distribution. Standards bodies and
operators increasingly require that these detectors meet formal resilience criteria, not only that they perform well on clean data, but that they absorb adversarial disturbances without catastrophic degradation, recover reliably after attack occurrence, and permit system-level certification from component-level evaluation. Existing resilience frameworks for cyber-physical systems (CPS) define absorb-recover-adapt trajectories at an architectural level and have been operationalised for infrastructure under physical disruption~\citep{Linkov2013, Cassottana2023}, but they do not treat ML-based anomaly detectors as first-class components: there is no formal connection between adversarial ML robustness constructs and the resilience curves, absorption capacities, and recovery trajectories that quantitative CPS resilience theory provides.
 
On the adversarial ML side, the gap is equally sharp. Adversarial robustness in ICS anomaly detection has attracted sustained research attention~\citep{Alotaibi2023, Anthi2021} and produced a well-established vocabulary of attack strategies and defences. Yet robustness is currently treated as a scalar property to be \emph{measured} on an individual detector, not a compositional property to be \emph{derived}, \emph{bounded}, or \emph{certified} across a deployed system. There is no formal language to ask how the robustness of a heterogeneous ICS network relates to the robustness of its constituent detection nodes, no dynamic model of how detection performance evolves as an attack persists or ceases, and no framework that connects benchmark results to the resilience standards bodies~\citep{NIST2023}.
 
We argue that both gaps share a common cause: adversarial robustness, as currently formalised, is a special and highly restrictive instantiation of a more general concept that engineering and ecological sciences have studied for decades under the name of \emph{system resilience}. The $\ell_p$-norm perturbation budget that defines the adversarial robustness threat model is a degenerate disturbance class that excludes the semantically structured, temporally extended, and physically constrained attacks characteristic of the actual ICS threat landscape. The certified radius is a single point on an absorption curve, not the full profile. The binary robust/not-robust
classification collapses a degradation function to a step function, a poor model for systems where graceful degradation under sustained attack matters more than sharp threshold behaviour. Resilience theory, by contrast, provides precisely the constructs that adversarial robustness lacks: disturbance class, absorption capacity, recovery
trajectory, and degradation function, as the literature reviewed in Section~\ref{sec:background} makes clear. The reframing this paper proposes is consequently structural, and it enables the compositional reasoning that ICS resilience certification requires.
 
This paper makes three main contributions. First, we provide a \textbf{formal mapping} of the core constructs of resilience theory, i.e. disturbance class, absorption capacity, recovery trajectory, and degradation function, onto the adversarial ML setting for OT anomaly detection, showing in each case that the standard adversarial robustness formalism captures only a degenerate special case of the corresponding resilience construct (Section~\ref{sec:formalism}).
Second, we derive a \textbf{compositional resilience bound}: we model a heterogeneous ICS network as a directed graph of anomaly detection nodes and show that the system-level resilience under a multi-point attack is bounded above by a function of the minimum node-level absorption capacity weighted by coupling strength along the attack path (Section~\ref{sec:composition}). This result demonstrates formally that the binding constraint on system-level resilience is the coupling-adjusted absorption capacity of each node along the attack path, so that the binding node need not coincide with the minimum-capacity node and its identification requires the network
coupling structure that node-level evaluation cannot observe.
Third, we introduce and empirically validate a set of \textbf{domain-appropriate resilience metrics} for OT anomaly detection, grounded in (a) extended  adversarial robustness  experiments based on methodology and architecture from our previous work~\citep{Stojanovic2022, Flatscher2024, Somma2024}, and (b) new adversarial attack and training experiments on the three-zone edge network conducted for this paper, showing that the metrics capture phenomena, in particular the absorption-degradation divergence under adversarial training, the campaign-level threat model for sustained attacks, and vulnerability amplification along coupled detection paths, that standard adversarial robustness benchmarks miss (Section~\ref{sec:empirical}).
 
The remainder of this paper is structured as follows.
Section~\ref{sec:background} surveys the empirical literature on adversarial robustness in OT intrusion detection, reviews resilience theory as applied to cyber-physical systems, and states the precise gap that motivates this work. Section~\ref{sec:formalism} formalises
the mapping between resilience-theoretic constructs and the adversarial ML setting. Section~\ref{sec:composition} develops the network model and derives the compositional resilience bound.
Section~\ref{sec:empirical} presents experimental validation.
Section~\ref{sec:design} discusses implications for IDS architecture,
certification standards, and open problems.
Section~\ref{sec:conclusion} concludes.

\section{Background and Related Work}
\label{sec:background}
 
This section establishes the theoretical and empirical context for the formalisation developed in Section~\ref{sec:formalism}. We organise the literature across three dimensions.
Section~\ref{subsec:bg_adv} surveys adversarial robustness research for ML-based IDS in OT and ICS environments, with emphasis on the limitations of current evaluation methodology.
Section~\ref{subsec:bg_resilience} reviews resilience theory as applied to cyber-physical systems, identifying the quantitative constructs, i.e. absorption, recovery trajectory, and degradation, that are relevant to the adversarial ML setting. Section~\ref{subsec:bg_gap} synthesises both bodies of work into a precise gap statement: to the best of our knowledge no existing work treats adversarial robustness as a compositional resilience property in a heterogeneous ICS anomaly detection context, and this is the main contribution the present paper makes.
 
\subsection{Adversarial Robustness in OT Intrusion Detection}
\label{subsec:bg_adv}
 
Machine learning (ML) and deep learning (DL) have become the dominant paradigm for anomaly-based intrusion detection in ICS and OT environments, offering detection capabilities for novel, zero-day attacks that rule-based systems miss~\citep{Bhamare2020, Stojanovic2020}.
However, the same properties that make ML detectors effective expose them to adversarial manipulation. Adversarial machine learning (AML) attacks craft inputs deliberately perturbed to cause misclassification, and their threat to IDS has been extensively documented~\citep{Alotaibi2023, NIST2023}.
 
The predominant attack model constrains perturbations by an $\ell_p$-norm budget $\eps$~\citep{Szegedy2014, Goodfellow2015, Madry2018, Biggio2018}.
Within this model, Projected Gradient Descent (PGD)~\citep{Madry2018} has emerged as the reference white-box attack.
PGD and the related Fast Gradient Sign Method (FGSM)~\citep{Goodfellow2015} substantially reduce classification accuracy of DNN-based IDS across standard benchmarks~\citep{Jmila2022, Alotaibi2023}. In the ICS domain, \citet{Anthi2021} demonstrated that Jacobian-based saliency map attacks reduce classifier performance by six to eleven percentage points on an
authenticated power system dataset; \citet{Tariq2022} showed that state-of-the-art time-series anomaly detectors, including DNN and GNN architectures on aerospace and CPS datasets, can be rendered ineffective under FGSM and PGD, with detection rates falling to near zero in the worst case. Adversarial perturbations of time-series inputs pose particular challenges because temporal correlations between consecutive sensor readings create dependencies that standard pointwise $\ell_p$ budgets do not capture~\citep{Fawaz2019, Zizzo2020}. 
Empirical results from OT testbed deployments that directly motivate the present paper are reported in~\citep{Flatscher2024, Somma2024}.
 
Defensive responses have broadly followed two paths. Adversarial training~\citep{Madry2018} augments the training set with perturbed examples, improving empirical robustness at the cost of reduced clean accuracy, a fundamental trade-off that is provably inherent and has
been documented in both the general AML literature~\citep{Tsipras2019} and for autoencoder-based anomaly detectors specifically~\citep{Beggel2019, Jmila2022}. Certified defences, notably randomised smoothing~\citep{Cohen2019}, provide provable guarantees that a classifier's prediction is constant within a certified $\ell_p$ ball, but remain computationally expensive, scale poorly to high-dimensional sensor time-series, and provide
guarantees only for $\ell_p$-bounded perturbations.
 
A consistent theme in the ICS-specific literature is the disconnect between benchmark robustness and operational behaviour. \citet{Mr2021} conducted experiments on an operational water treatment plant testbed and concluded that threat models must extend beyond gradient attacks to encompass stealthy attacks in which an adversary gradually manipulates
sensor readings to accumulate bias below detection thresholds. In ICS deployments specifically, adversarial perturbations must additionally satisfy physical feasibility constraints: \citet{Jia2021} demonstrate that effective evasion in CPS requires simultaneously deceiving the neural-network anomaly detector \emph{and} the physics-based invariant checker enforcing process constraints -- a combination that no fixed $\ell_p$-norm budget can model. \citet{Sheikh2023} further document that adversarial defence strategies
for ML-based CPS security must account for these domain-specific constraints to provide meaningful protection. Such slow-drift or integrity attacks do not correspond to any fixed $\ell_p$-ball perturbation and therefore lie outside the scope of both standard adversarial training and certified defences. The NIST taxonomy of adversarial ML~\citep{NIST2023} further underlines this gap: it categorises attacks by threat model and goal but provides no
compositional or system-level robustness framework. The practical implication is that ICS operators cannot derive system-level robustness assurances from component-level benchmark results.
 
\subsection{Resilience Theory in Cybersecurity and Cyber-Physical Systems}
\label{subsec:bg_resilience}
 
Resilience has been studied as a systems property since Holling's foundational work on ecological stability~\citep{Holling1973}. Engineering resilience theory reframes this concept around four sequential phases: \emph{plan/prepare}, \emph{absorb}, \emph{recover},
and \emph{adapt}~\citep{Linkov2013, NRC2012}, and has been formalised as the Resilience Matrix~\citep{Linkov2013}, operationalised across military, infrastructure, and ecological
contexts~\citep{Linkov2019}, and applied to quantitative assessment of CPS under disruption~\citep{Cassottana2023, Perrett2023}.
 
The distinction between engineering and ecological resilience is consequential here. Engineering resilience emphasises return time to a prior steady state~\citep{Pimm1984}; ecological resilience emphasises the magnitude of disturbance a system can absorb before transitioning to an alternative stable state~\citep{Holling1973}. Bruneau et al.~\citep{Bruneau2003} organise infrastructure seismic resilience around four properties, i.e. robustness, redundancy, resourcefulness, and rapidity, widely adopted in subsequent
frameworks~\citep{Linkov2017}.
 
In the CPS and cybersecurity domain, \citet{Segovia2024} define cyber-resilience as the ability to prepare, absorb, recover, and adapt to adverse cyber effects, identifying network cascade failures as a principal threat to absorption and recovery phases: a network
susceptible to cascading failures has impaired absorption capacity regardless of individual component robustness. This observation is directly relevant to heterogeneous ICS deployments.
\citet{Cassottana2023} systematise quantitative CPS resilience assessment as a three-step process, i.e. CPS description, disruption scenario identification, and performance evaluation, noting that existing approaches treat cyber and physical layers in relative
isolation. \citet{Perrett2023} argue that OT environments have inherited the vocabulary of information security resilience without the process-safety constraint logic required in safety-critical settings, and introduce a hierarchical trajectory model distinguishing
engineering resistance, organisational recovery, and adaptive capacity. Functional safety standards for programmable electronic systems, notably IEC~61508~\citep{IEC61508}, specify quantitative Safety Integrity Level (SIL) requirements for safety-critical components; how
ML-based anomaly detectors relate to these SIL requirements has not been formally addressed, constituting a further motivation for the formalisation developed here. 
 
Quantitative resilience metrics for CPS have been proposed by several groups. \citet{Linkov2013} propose performance-versus-time curves capturing both degradation depth and recovery speed, with resilience expressed as the area between the nominal and post-disturbance trajectory. \citet{Henry2012} provide an axiomatic, time-dependent treatment of system resilience as a measurable engineering quantity, and \citet{Hosseini2016} survey formal definitions and measures of system resilience across the engineering and infrastructure literature.
\citet{Haque2021} survey formal CPS security frameworks and propose quantitative resilience analytics. An integrated framework combining Markov-chain cyber-anomaly modelling with continuous degradation and recovery modelling in the physical domain is developed
by~\citet{Li2025}, demonstrating that resilience curves provide richer diagnostic information than scalar robustness metrics.
 
Despite this body of work, existing quantitative resilience frameworks for CPS share a common limitation: they are defined at the system or subsystem level, not at the level of an individual ML-based anomaly detection component, and they do not characterise how perturbation-specific properties of ML detectors, i.e. adversarial vulnerability, certified radius, and detection threshold, relate to the constituent phases of the resilience curve. The connection between adversarial ML robustness and engineering resilience constructs has not been formalised.
 
\subsection{The Gap: Robustness as Compositional Resilience}
\label{subsec:bg_gap}
 
Table~\ref{tab:related_work_gap} organises the surveyed literature against the four dimensions on which the gap is claimed. Each row covers one representative work; the columns correspond to the four properties that a complete compositional resilience theory for ICS
anomaly detection requires. The table makes the structural gap precise: no existing work achieves contribution (\checkmark) in all four columns simultaneously.
 
\begin{table}[t]
    \centering
    \footnotesize
    \caption{Structured comparison of key related work across the four dimensions of the claimed gap. \checkmark~= fully addressed; $\sim$~= partially addressed; {---}~= not addressed.
    Column~(a): ML-based anomaly detectors treated as first-class components with explicitly modelled adversarial properties.
    Column~(b): formal adversarial ML threat model.
    Column~(c): temporal or dynamic formulation of detector behaviour.
    Column~(d): compositional analysis across a \emph{network} of heterogeneous ML-based detectors, with coupling constraints.
    Column~(e): role of each work relative to the present paper's contribution.}
    \label{tab:related_work_gap}
    \renewcommand{\arraystretch}{1.35}
    \begin{tabular}{@{}p{3.3cm} c c c c p{9.0cm}@{}}
        \toprule
        \textbf{Reference} &
        \textbf{(a)} &
        \textbf{(b)} &
        \textbf{(c)} &
        \textbf{(d)} &
        \textbf{(e) Role in context} \\
        \midrule
        \multicolumn{6}{@{}l}{\textit{Adversarial ML}} \\[2pt]
 
        \citet{Madry2018}
          & \checkmark & \checkmark & {---} & {---}
          & Defines PGD; operationalises $\ell_\infty$ evasion budget used in all experiments. \\
 
        \citet{Cohen2019}
          & \checkmark & \checkmark & {---} & {---}
          &  
          Certified radius can be interpreted as a static, pointwise lower bound on a systems perturbation absorption capacity under a fixed $\ell_2$ threat model. \\
 
        \citet{Alotaibi2023}
          & \checkmark & \checkmark & {---} & {---}
          & Surveys adversarial evasion against ML-based IDS in OT;
            no system-level or compositional analysis. \\
 
        \citet{Tariq2022}
          & \checkmark & \checkmark & $\sim$ & {---}
          & Temporal adversarial anomaly detection in OT; temporal structure is in the data, not formalised as a trajectory. \\
 
        \citet{NIST2023}
          & \checkmark & \checkmark & {---} & {---}
          & Adversarial ML threat taxonomy and guidance;
            no system-level resilience or composition model. \\
 
        \citet{Flatscher2024}
          & \checkmark & \checkmark & {---} & {---}
          & Our initial work -- investigates HoE for AE-based IDS and provides the $\hat{\mathcal{A}}$ estimator reused in Section~5.2. 
          \\
 
        \midrule
        \multicolumn{6}{@{}l}{\textit{Resilience theory for CPS}} \\[2pt]
 
        \citet{Bruneau2003}
          & {---} & {---} & \checkmark & $\sim$
          & Defines the resilience quadrangle; primary source of the
            absorb--recover--adapt constructs formalised here. \\
 
        \citet{Linkov2013}
          & {---} & {---} & \checkmark & $\sim$
          & Absorb--recover--adapt framework for infrastructure;
            applied to physical systems, not ML detectors. \\
 
        \citet{Cassottana2023}
          & {---} & {---} & \checkmark & $\sim$
          & Reviews CPS resilience models and methods; no adversarial ML component or detector-level formalisation. \\
 
        \citet{Segovia2024}
          & {---} & {---} & \checkmark & $\sim$
          & Cyber-resilience and cascade failure survey for CPS; closest to column~(d) but uses probabilistic system states, not adversarial ML constructs. \\
 
        \citet{Li2025}
          & $\sim$ & {---} & \checkmark & $\sim$
          & Markov process model of cyber/security states with physical degradation curves; probabilistic, not worst-case adversarial; no compositional bound across ML detectors. \\
 
        \midrule
        \textbf{This paper}
          & \checkmark & \checkmark & \checkmark & \checkmark
          & First work to derive a compositional adversarial resilience bound for heterogeneous ICS ML detector networks. \\
        \bottomrule
    \end{tabular}
\end{table}
 
On the adversarial ML side (upper block), every work addresses columns~(a) and~(b) by construction, but columns~(c) and~(d) are uniformly absent. Robustness is evaluated as a scalar property of a single detector at a single point in time; there is no temporal model
of how detection performance evolves as an attack persists or ceases, and no analysis of how robustness propagates across a network of coupled detectors. \citet{Tariq2022} receives a partial mark on column~(c) because their experiments use time-series input data and
observe degradation across sustained attacks, but they do not formalise a recovery trajectory or define a time-indexed performance function: the temporal structure is in the data, not in the evaluation framework.
 
On the resilience theory side (lower block), columns~(c) and~(d) are addressed at the level of physical systems and CPS architectures, but columns~(a) and~(b) are absent. The two most recent and relevant papers warrant individual examination, as they represent the closest
existing approaches. \citet{Segovia2024} survey cyber-resilience for CPS and explicitly address cascade failure analysis, which is the closest structural analogue to the compositional resilience result in Eq.~\eqref{eq:comp_bound}; column~(d) therefore receives~$\sim$ rather than~{---}. However, their cascade analysis operates on system
states and probabilistic failure propagation, not on the adversarial ML constructs, i.e. disturbance class, absorption capacity, certified radius, that characterise individual ML-based detector behaviour. The paper does not derive a bound relating component-level detector
properties to system-level resilience under adversarial attack, and does not treat detectors as components with explicitly modelled evasion vulnerability. \citet{Li2025} develop a quantitative CPS resilience framework integrating Markov-chain cyber-anomaly modelling
with continuous physical degradation curves, and demonstrate that resilience curves provide richer diagnostic information than scalar robustness metrics -- a result consistent with the present paper's findings on the absorption--degradation divergence. Their anomaly
component is a Markov-chain state machine, not an ML-based anomaly detector with an adversarial threat model, and their framework is applied at the system level without a compositional result connecting component-level properties to the system-level curve. Column~(a) receives~$\sim$ because ML anomaly detection is discussed but not modelled with adversarial ML constructs; column~(b) remains~{---} because their disruption model is probabilistic, not worst-case adversarial.
 
The gap in column~(d) across both blocks is therefore precise: no existing work provides a \emph{formal, adversarially-grounded compositional bound} relating the adversarial properties of individual ML-based anomaly detectors to system-level resilience under a
multi-point attack on a coupled detection network. 
The adversarial ML literature lacks the compositional apparatus; the resilience literature lacks the adversarial ML threat model. The contribution of this paper is to bridge both directions simultaneously, providing the formal constructs
(Sections~\ref{subsec:disturbance}--\ref{subsec:degradation}) and the compositional result (Eq.~\eqref{eq:comp_bound}) that neither body of work contains.

\section{Formalising the Connection: Robustness as Resilience}
\label{sec:formalism}
 
We adopt the constructs of \emph{disturbance class}, \emph{absorption capacity}, \emph{recovery trajectory}, and \emph{degradation function} from engineering resilience theory~\citep{Linkov2013, Bruneau2003} and provide for the first time formal definitions of each within the adversarial ML
setting for OT anomaly detection. 
Throughout this section we model an anomaly detection system as a deterministic binary classifier $f : \mathcal{X} \to \{0,1\}$ operating over an observation space $\mathcal{X} \subseteq \R^d$, with a performance metric $\Phi : \mathcal{X} \to [0,1]$ that maps
a sequence of observations to a scalar detection performance value (e.g., true positive rate or $F_1$ score).  Additionally, it should be noted that throughout this section and Section~\ref{sec:composition}, $\|\cdot\|_p$ denotes the $\ell_p$-norm for a fixed
$p \in [1, \infty]$ -- the choice of $p$ is a modelling decision that does not affect the  reasoning and results. In the empirical instantiation of Section~\ref{sec:empirical} we fix $p = 2$, consistent with the unclipped minimum-norm PGD attack used to
estimate the Hardness of Evasion in Eq.~\eqref{eq:absorb_estimate}.
The four constructs and their standard adversarial ML analogues are summarised in Table~\ref{tab:mapping}; the remainder of this section develops each in turn.
 
\subsection{Disturbance Class}
\label{subsec:disturbance}
 
In resilience theory, resilience is always defined \emph{relative to a specified disturbance set}~\citep{Linkov2013, Holling1973}: a system is not resilient in the abstract, but resilient with respect to a particular class of disruptions. The adversarial ML literature
implicitly adopts this principle but imposes a disturbance class so restrictive that it excludes the most operationally relevant attacks in ICS environments. This section defines the \emph{disturbance class}, i.e. what is considered as a valid attack in an ICS  environment. It justifies that an attack is not just a single number, but a sequence of manipulations over time, and that sequence must satisfy certain rules simultaneously.
 
\paragraph{Disturbance class for anomaly detection.}
Let $\mathcal{X} \subseteq \R^d$ be the observation space of an anomaly detector and $T \in \mathbb{N}$ a temporal horizon. We model the adversarial environment as a \emph{disturbance class} $\dist$, defined as a set of perturbation sequences $\boldsymbol{\delta} = (\delta_1, \ldots, \delta_T)$ with $\delta_t \in \R^d$, characterised by three structural conditions:
\begin{enumerate}[label=(\roman*)]
    \item \textbf{Budget constraint:}
          $B(\boldsymbol{\delta}) \leq b$ for some budget function $B : (\R^d)^T \to \R_{\geq 0}$ and scalar $b \geq 0$, where $b$ controls the total permitted attack magnitude;
    \item \textbf{Semantic constraint:}
          $\delta_t \in \mathcal{C}(x_t)$ for all $t$, where $\mathcal{C}(x_t) \subseteq \R^d$ is a context-dependent feasibility set encoding physical plausibility conditions on observation $x_t$;
    \item \textbf{Temporal structure:}
          $\boldsymbol{\delta}$ belongs to a prescribed trajectory family $\mathcal{T} \subseteq ((\R^d)^T)$, which may encode correlation or monotonicity constraints across time.
\end{enumerate}
This three-part specification -- budget, semantics, temporal structure, characterises an attack as more than a single perturbation magnitude: it is a structured sequence subject to operational constraints that must hold simultaneously.
 
The standard adversarial ML threat model corresponds to the \emph{degenerate} special case $T = 1$, $B(\delta_1) = \norm{\delta_1}_p$, $\mathcal{C}(x_1) = \R^d$, and $\mathcal{T} = \{\delta_1\}$, yielding the familiar $\eps$-ball:
\begin{equation}
    \dist_{\eps} = \bigl\{\delta \in \R^d :
    \norm{\delta}_p \leq \eps \bigr\}.
    \label{eq:lp_ball}
\end{equation}
This degenerate class is point-in-time, unconstrained by physical semantics, and admits no temporal structure. All three restrictions are consequential in OT environments.
 
\paragraph{Semantic constraints in ICS.}
ICS sensor readings are constrained by physical laws: flow rates, pressures, and temperatures evolve according to underlying process dynamics that bound both the magnitude and the rate of change of any plausible sensor value. A perturbation that exceeds physical process bounds is immediately detectable by process-invariant monitors~\citep{Mr2021}, and an attacker operating under the constraint of long-term evasion will respect these bounds. The $\ell_p$-ball $\dist_\eps$ ignores this: it places equal weight on all directions in $\R^d$, including physically impossible ones. Formally, the feasibility set $\mathcal{C}(x_t)$ for a sensor reading $x_t^{(i)}$ of channel $i$ at time $t$ is:
\begin{equation}
    \mathcal{C}^{(i)}(x_t) =
    \bigl\{\delta^{(i)} :
    x_t^{(i)} + \delta^{(i)} \in [l^{(i)}, u^{(i)}],\;
    |\delta^{(i)}| \leq r^{(i)} \bigr\},
    \label{eq:semantic_constraint}
\end{equation}
where $[l^{(i)}, u^{(i)}]$ is the physically admissible range for channel $i$ and $r^{(i)}$ is a per-step rate-of-change bound derived from process dynamics. The full disturbance class for semantically constrained ICS attacks is then
$\dist_{\mathrm{ICS}} = \{\boldsymbol{\delta} : \delta_t \in
\prod_i \mathcal{C}^{(i)}(x_t),\; t = 1,\ldots,T\}$.
 
\paragraph{Temporal structure: slow-drift and staged attacks.}
A slow-drift attack~\citep{Mr2021} introduces perturbations that are individually small at each time step but accumulate a significant bias over a horizon $T$. This is captured by choosing $B(\boldsymbol{\delta}) = \norm{\sum_{t=1}^T \delta_t}_p$ (cumulative budget) rather than $\norm{\delta_1}_p$ (single-step budget), and a trajectory family $\mathcal{T}$ that imposes monotonicity: $\delta_{t+1}^{(i)} \geq \delta_t^{(i)}$ for channels targeted by
the attack. A single-step $\ell_p$-ball cannot represent either constraint. Similarly, a staged spoofing attack that selectively corrupts one subsystem before propagating to a second lies entirely outside $\dist_\eps$, which does not model temporal sequencing.

\subsection{Absorption Capacity}
\label{subsec:absorption}
 
In resilience theory, absorption capacity is the degree to which a
system can sustain performance under disturbance without structural
change~\citep{Bruneau2003, Linkov2013}. We formalise this for an
anomaly detector as the supremum of the disturbance magnitude at
which detection performance remains above a prescribed threshold.
 
\paragraph{Absorption capacity and absorption curve.}
We characterise the resilience of an anomaly detector $f : \mathcal{X} \to \{0,1\}$ under a disturbance class $\dist$ parameterised by budget $b$, at performance threshold $\tau \in (0,1]$, through two related quantities. The first is the \emph{absorption capacity}: the largest budget at which worst-case detection performance still meets the threshold,
\begin{equation}
    \abs(f, \dist, \tau) \;=\;
    \sup \bigl\{ b \geq 0 :
    \inf_{\boldsymbol{\delta} \in \dist(b)}
    \Phi\bigl(f,\, x + \boldsymbol{\delta}\bigr) \geq \tau
    \bigr\},
    \label{eq:absorption}
\end{equation}
where $\dist(b)$ denotes the disturbance class at budget level $b$. The second is the \emph{absorption curve}: the same worst-case performance, viewed as a function of the budget rather than evaluated at a single threshold,
\begin{equation}
    A_f(b) \;\triangleq\;
    \inf_{\boldsymbol{\delta} \in \dist(b)}
    \Phi\bigl(f,\, x + \boldsymbol{\delta}\bigr),
    \qquad b \geq 0 ,
    \label{eq:absorption_curve}
\end{equation}
recording the lowest detection performance attainable by any disturbance within budget $b$.
 
\begin{figure}
    \centering
    \includegraphics[width=0.7\linewidth]{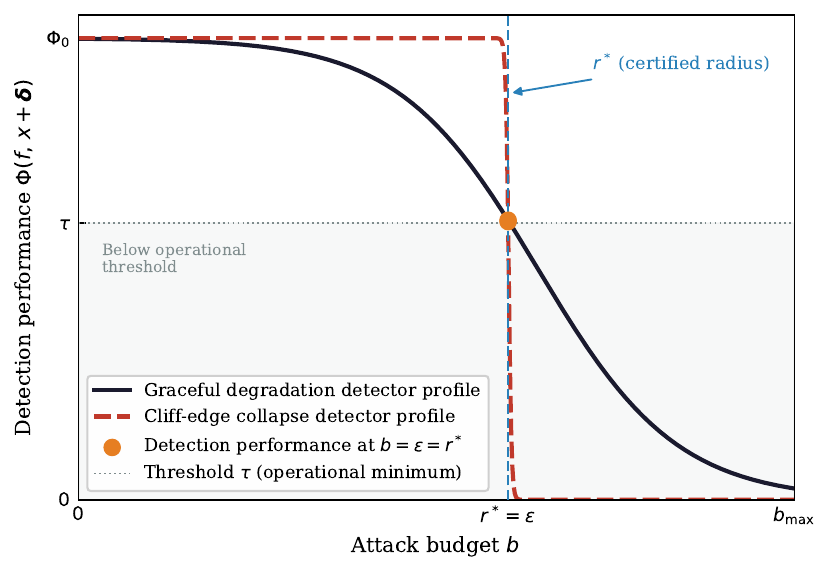}
    \caption{Schematic absorption curves for two detector profiles under disturbance class $\mathcal{D}$. The graceful degradation profile (solid) represents the general case of a smooth, continuous degradation (Eq.~\eqref{eq:degradation}) that retains partial detection capability beyond the absorption threshold. The cliff-edge collapse profile (dashed) represents the special step-function case, which is the implicit assumption of the binary robust/not-robust evaluation framework. Detection performance $\Phi$ is
    plotted against attack budget $b$. The certified radius $r^*$
    (vertical dashed line) and accuracy under PGD at fixed $\eps$
    (single marker) are both point evaluations on this curve.
    The full profile captures graceful degradation, which neither
    metric reflects.}
    \label{fig:absorption_curve}
\end{figure}

\noindent The curve $A_f$ and the capacity $\abs(f, \dist, \tau)$
describe the same disturbance response on two different axes, and the
distinction is used throughout this paper. The curve is
\emph{performance-valued}: its outputs are values of $\Phi$ on the
performance axis, and it plots worst-case detection performance
against attack budget $b$ (Figure~\ref{fig:absorption_curve}). The
capacity is \emph{budget-valued}: a single value of $b$ on the budget
axis. The two are linked by reading the curve at the threshold $\tau$,
$\abs(f, \dist, \tau) = \sup\{b \geq 0 : A_f(b) \geq \tau\}$ -- the
largest attack budget at which the curve remains at or above $\tau$.
 
\paragraph{Standard metrics as point evaluations.}
The two primary formal robustness metrics in adversarial ML are
both degenerate point evaluations on this absorption curve.
 
\begin{itemize}
    \item \textbf{Certified radius}~\citep{Cohen2019}: under
          randomised smoothing, the certified radius $r^*$ is the
          largest $\ell_2$-ball radius within which the smoothed
          classifier $g$ is guaranteed to return a fixed label.
          This corresponds to $\abs(g, \dist_\eps^{(\ell_2)}, \tau)$
          evaluated at a single threshold $\tau$ equal to the
          majority-vote confidence of $g$ at the nominal input,
          a scalar extracted from a single point on the absorption
          curve under a single disturbance class.
 
    \item \textbf{Accuracy under PGD}~\citep{Madry2018}: the
          fraction of inputs correctly classified after PGD attack
          at budget $\eps$ corresponds to $\Phi(f, x + \delta^*)$
          at a fixed $\eps$, averaged over the input distribution.
          This is a point evaluation of the absorption curve at
          $b = \eps$ under the $\ell_\infty$-ball disturbance class.
\end{itemize}
 
Both metrics fix $b$ and read off $\Phi$; neither traces the full
curve $b \mapsto \Phi(f, x+\delta^*(b))$. As a consequence, two
detectors that achieve identical accuracy under PGD at budget $\eps$
may have radically different absorption curves: one may maintain
near-nominal performance up to $b = \eps$ and then collapse sharply,
while the other may degrade smoothly from $b = 0$ (Figure~\ref{fig:absorption_curve}). The distinction
is operationally critical in ICS contexts, where graceful degradation
under partial compromise is more valuable than binary pass/fail at a
single threshold (Section~\ref{subsec:degradation}).
 
\subsection{Recovery Trajectory}
\label{subsec:recovery}
 
Standard adversarial robustness evaluation is a \emph{static} measurement: the detector is evaluated at a single point in time under a worst-case perturbation, and the resulting scalar is taken as the measure of robustness. This framing has no temporal dimension. Resilience theory, by contrast, treats recovery as a first-class property: a system that absorbs a disturbance poorly but recovers rapidly may be more operationally acceptable than one that resists the disturbance initially but degrades permanently once breached~\citep{Bruneau2003, Linkov2013}. We introduce a dynamic formulation that brings this dimension into the adversarial ML setting for OT anomaly detection.
 
\paragraph{Recovery trajectory.}
Let $f$ be an anomaly detector and $\{x_t\}_{t \geq 0}$ a time-indexed sequence of sensor observations. Let $t_0$ denote the onset time of an attack $\boldsymbol{\delta} \in \dist$, so that the observed sequence under attack is $\tilde{x}_t = x_t + \delta_t$ for $t \in [t_0, t_0 + T]$ and $\tilde{x}_t = x_t$ otherwise. We characterise the post-attack behaviour of $f$ by its \emph{recovery trajectory},
\begin{equation}
    \recov(f, t_0, \dist) \;:\; t \;\mapsto\;
    \Phi\bigl(f,\, \tilde{x}_t\bigr),
    \quad t \geq t_0,
    \label{eq:recovery_traj}
\end{equation}
where $\Phi(f, \tilde{x}_t)$ is detection performance evaluated over a sliding window ending at time $t$.
 
Two scalar summaries of the recovery trajectory are operationally relevant:
 
\begin{itemize}
    \item \textbf{Recovery time} $T_r$: the smallest $t > t_0 + T$ such that $\Phi(f, \tilde{x}_t) \geq \tau$ for all subsequent $t' \geq t$, where $\tau$ is the nominal          operating threshold. Formally:
          \begin{equation}
              T_r = \inf \bigl\{t > t_0 + T :
              \Phi(f, \tilde{x}_{t'}) \geq \tau\;
              \forall\, t' \geq t \bigr\} - (t_0 + T).
              \label{eq:recovery_time}
          \end{equation}
 
    \item \textbf{Recovery completeness} $\omega_r \in [0,1]$: the asymptotic performance level relative to the pre-attack baseline:
          \begin{equation}
              \omega_r = \frac{
              \lim_{t \to \infty} \Phi(f, \tilde{x}_t)
              }{
              \Phi(f, x_t)\big|_{t < t_0}
              }.
              \label{eq:recovery_completeness}
          \end{equation}
          $\omega_r = 1$ indicates full recovery;
          $\omega_r < 1$ indicates permanent degradation.
\end{itemize}
 
\paragraph{Absence in standard adversarial ML.}
The adversarial robustness literature has no counterpart to the recovery trajectory. Existing evaluation protocols sample worst-case performance at a fixed perturbation magnitude; they
do not track detector behaviour as the attack evolves, ceases, or transitions between phases. This absence is not merely a gap in benchmarking practice: it reflects a structural assumption that the attack is instantaneous and the detector stateless. In OT environments, neither assumption holds. ICS anomaly detectors typically operate over sliding windows of sensor time-series, and real attacks, including staged spoofing and slow-drift attacks, unfold over minutes or hours~\citep{Mr2021}. The recovery trajectory is therefore not a theoretical refinement but a practical necessity for assessing detector behaviour in the deployment context the paper addresses. Its empirical estimation from OT testbed data is demonstrated in Section~\ref{subsec:profiles}.
 
\subsection{Degradation Function}
\label{subsec:degradation}
 
The final construct concerns the shape of the transition between
nominal performance and fully compromised performance as attack
intensity increases. In resilience theory this transition is
modelled explicitly as a continuous degradation function; in
adversarial ML it is collapsed, by convention, to a binary
threshold.
 
\paragraph{Degradation function.}
The shape of the transition between nominal and compromised performance is captured by the \emph{degradation function} of detector $f$ under disturbance class $\dist$,
\begin{equation}
    \degrad_f : b \;\mapsto\;
    1 - \frac{
    \inf_{\boldsymbol{\delta} \in \dist(b)}
    \Phi(f,\, x + \boldsymbol{\delta})
    }{
    \Phi(f, x)
    },
    \quad b \geq 0,
    \label{eq:degradation}
\end{equation}
where $\Phi(f, x)$ is the nominal (unperturbed) detection performance. By construction $\degrad_f(0) = 0$ (no degradation at zero budget), $\degrad_f$ is non-decreasing in $b$, and $\degrad_f(b) \in [0,1]$ for all $b$.
 
\noindent The degradation function $\degrad_f$ is the normalised
complement of the absorption curve $A_f$
(Eq.~\eqref{eq:absorption_curve},
Eq.~\eqref{eq:absorption_curve}). The two are related by
\begin{equation}
    \degrad_f(b) \;=\; 1 \;-\; \frac{A_f(b)}{\Phi(f, x)},
    \qquad b \geq 0,
    \label{eq:degrad_absorb_rel}
\end{equation}
so that $A_f$ records what the detector retains and $\degrad_f$
records the fraction it has lost. The absorption \emph{curve}
$A_f$ is performance-valued (it returns values of $\Phi$, in
$[0, 1]$ on the performance axis); the absorption \emph{capacity}
$\abs(f, \dist, \tau)$ of Eq.~\eqref{eq:absorption} is by
contrast budget-valued and is recovered from $A_f$ as
\begin{equation}
    \abs(f, \dist, \tau) \;=\; \sup\bigl\{b \geq 0 :
    A_f(b) \geq \tau \bigr\},
    \label{eq:abs_from_curve}
\end{equation}
i.e.\ the largest attack budget at which the curve remains at
or above the performance threshold $\tau$ -- equivalently, the
budget at which $\degrad_f(b) = 1 - \tau/\Phi(f, x)$. The
absorption curve and absorption capacity therefore live on
different axes (curve: performance axis; capacity: budget axis)
and should not be conflated, even though both are labelled by
$\tau$ and parameterised through the same disturbance class
$\dist$.
 
\paragraph{The standard binary framing as a degenerate case.}
The conventional adversarial robustness classification -- robust
if $\Phi(f, x + \delta^*) \geq \tau$, not robust otherwise,
corresponds to the indicator function:
\begin{equation}
    \degrad_f^{\mathrm{bin}}(b) \;=\;
    \mathbf{1}\bigl[b > b^*\bigr],
    \label{eq:step_degradation}
\end{equation}
where $b^* = \abs(f, \dist, \tau)$ is the absorption capacity at
threshold $\tau$. This is a step function: zero degradation for all
$b \leq b^*$, full degradation for all $b > b^*$. The step function
is a special case of Eq.~\eqref{eq:degradation} only when
the detector transitions instantaneously from nominal to zero
performance at a sharp threshold, a model that is empirically
false for virtually all ML-based detectors and operationally
misleading for ICS deployment.
 
\paragraph{Smooth degradation in OT contexts.}
In practice, the degradation function of an ML-based anomaly
detector is continuous and increasing, with the rate of increase
dependent on both the detector architecture and the disturbance
class. For example, under a slow-drift attack, $\degrad_f(b)$
increases gradually as the cumulative perturbation budget $b$
grows, reflecting the progressive shift of sensor readings toward
the anomaly-free region of the detector's decision boundary. An
operator calibrating a detection threshold in an OT environment
needs to know the slope and inflection point of $\degrad_f$,
not merely the location of $b^*$. For certification purposes,
a detector with a shallow $\degrad_f$ (graceful degradation) is
preferable to one with the same $b^*$ but a steep $\degrad_f$
(cliff-edge collapse), even though both would receive identical
scores under standard adversarial robustness benchmarks. The
estimation of $\degrad_f$ from experimental data is addressed
in Section~\ref{subsec:profiles}.
 
 
Table~\ref{tab:mapping} summarises the four mappings established in this section. In each row, the resilience construct in the left column subsumes the adversarial ML analogue in the centre column; the standard metric in the right column is a point evaluation or degenerate special case of the corresponding formal construct.
 
\begin{table}[t]
    \centering
    \caption{Mapping of resilience-theory constructs
    onto the adversarial ML setting for OT anomaly detection.
    Each standard metric is a degenerate special case of the corresponding resilience construct.}
    \label{tab:mapping}
    \renewcommand{\arraystretch}{1.35}
    \begin{tabular}{@{}p{4cm} p{5.0cm} p{5cm}@{}}
        \toprule
        \textbf{Resilience} \newline \textbf{Construct} &
        \textbf{Adversarial ML Analogue} \newline \textbf{(this paper)} &
        \textbf{Standard Metric} \newline \textbf{(degenerate case)} \\
        \midrule
        Disturbance class \newline
        (Sec.~\ref{subsec:disturbance}) &
        $\ell_p$-ball perturbation \newline model $\dist_\eps$ &
        $\ell_p$-norm bound $\eps$ \\[4pt]
 
        Absorption capacity \newline
        (Sec.~\ref{subsec:absorption}) &
        Detection performance \newline under worst-case attack &
        Certified radius $r^*$; \newline accuracy under PGD \\[4pt]
 
        Recovery trajectory \newline
        (Sec.~\ref{subsec:recovery}) &
        Post-attack performance \newline dynamics of detector $f$ &
        -- \emph{(no analogue in} \newline
            \emph{standard AML evaluation)} \\[4pt]
 
        Degradation function \newline
        (Sec.~\ref{subsec:degradation}) &
        Continuous performance \newline loss as $b$ increases &
        Binary robust / not \newline robust at threshold $b^*$ \\
        \bottomrule
    \end{tabular}
\end{table}

\section{Compositional Resilience in Heterogeneous ICS Networks}
\label{sec:composition}
 
The constructs formalised in Section~\ref{sec:formalism} characterise the resilience of an \emph{individual} anomaly detection node in isolation. In practice, ICS and OT deployments are heterogeneous: a production network will host multiple detection components, e.g. network intrusion detection, process anomaly monitors and historian integrity checkers, whose inputs and alarm outputs are coupled through data flows and event propagation paths. The system-level resilience of such a network is not simply the minimum of its node-level resilience profiles but it depends on the coupling topology. This section develops the formal apparatus to make that claim precise. Section~\ref{subsec:network_model} introduces the ICS network model. Section~\ref{subsec:bound} states and proves the main theoretical result. Section~\ref{subsec:eval_implications} draws the implication for adversarial evaluation practice and illustrates the compositional vulnerability with a constructed three-node example.

\subsection{Network Model}
\label{subsec:network_model}
 
\paragraph{Network model.}
We model an \emph{ICS detection network} as a directed weighted graph $G = (V, E, w)$, where:
\begin{itemize}
    \item $V = \{f_1, \ldots, f_n\}$ is a finite set of anomaly detection nodes, each $f_i : \mathcal{X}_i \to \{0,1\}$ operating over its local observation space $\mathcal{X}_i \subseteq \R^{d_i}$;
    \item $E \subseteq V \times V$ is a set of directed edges, where $(f_i, f_j) \in E$ indicates that the output or internal state of $f_i$ influences the input of $f_j$ (e.g., via shared sensor feeds, alarm aggregation, or data historian dependencies); and
    \item $w : E \to (0, 1]$ is a \emph{coupling strength} function, with $w_{ij} \triangleq w(f_i, f_j)$ denoting the coupling strength of edge $(f_i, f_j)$.
\end{itemize}
 
\paragraph{Coupling strength.}
The coupling weight $w_{ij} \in (0,1]$ quantifies the maximum fraction
of an adversarial disturbance targeting $f_i$ that can propagate, as an
induced disturbance on $f_j$, through the data dependency encoded by
edge $(f_i, f_j)$. Intuitively, it measures how much of an attack
mounted at one node reaches its neighbours: a value near $1$ passes the
disturbance through almost undiminished, whereas a small value means the
edge strongly attenuates it. Formally, let $\dist_i(b)$ denote the
disturbance class at $f_i$ with budget $b$, and let
$\iota_{ij} : \mathcal{X}_i \times \R^{d_i} \to \R^{d_j}$ be the
signal-propagation map from $f_i$ to $f_j$; we define
\begin{equation}
    w_{ij} \;=\; \sup_{\substack{\boldsymbol{\delta} \in \dist_i(b),\\
    b > 0}}
    \frac{\norm{\iota_{ij}(x_i,\, \boldsymbol{\delta})}_p}{b},
    \label{eq:coupling_weight}
\end{equation}
where the supremum is taken over all non-zero budgets $b$ and all
disturbances $\boldsymbol{\delta} \in \dist_i(b)$, and the norm is the
same $\ell_p$ norm used in the budget constraint of
Section~\ref{subsec:disturbance}. Dividing by $b$ gives $w_{ij}$
a precise Lipschitz-constant interpretation, it is the largest gain
$\iota_{ij}$ applies to an admissible disturbance, and in particular
$w_{ij} \leq 1$ whenever $\iota_{ij}$ is a contraction, as it is for the
signal-flow maps that arise in standard ICS architectures (shared sensor
aggregation, data historians, alarm forwarding).
 
\paragraph{Attack paths.}
A \emph{directed attack path} in $G$ is a sequence $\pi = (f_{i_1}, f_{i_2}, \ldots, f_{i_k})$ such that $(f_{i_l}, f_{i_{l+1}}) \in E$ for all $l = 1, \ldots, k-1$.
We denote by $\Pi(G)$ the set of all directed attack paths in $G$.
For a path $\pi$ and node $f_{i_j} \in \pi$, we define the \emph{cumulative coupling} reaching $f_{i_j}$ along $\pi$ as:
\begin{equation}
    W(\pi, j) \;=\;
    \prod_{l=1}^{j-1} w_{i_l i_{l+1}},
    \label{eq:cumulative_coupling}
\end{equation}
with the convention $W(\pi, 1) = 1$ (empty product). Since each $w_{i_l i_{l+1}} \leq 1$, the sequence $W(\pi, 1), W(\pi, 2), \ldots, W(\pi, k)$ is non-increasing: attack energy attenuates as it propagates along the path.

\subsection{Compositional Resilience Bound}
\label{subsec:bound}
 
We define the \emph{system-level resilience} along attack path $\pi$ as the normalised detection performance of the weakest node under a coordinated multi-point adversarial attack that distributes a total budget $b$ across the path, with each downstream node receiving attenuated budget proportional to the cumulative coupling:
\begin{equation}
    \varrho_{\mathrm{sys}}(\pi, b) \;=\;
    \min_{j = 1,\ldots,k}\;
    \Phi\!\left(f_{i_j},\;
    x_{i_j} + \boldsymbol{\delta}^*_{i_j}(b \cdot W(\pi,j))
    \right),
    \label{eq:sys_resilience}
\end{equation}
where $\boldsymbol{\delta}^*_{i_j}(b')$ is the worst-case disturbance
in $\dist_{i_j}(b')$ and $\Phi$ is the performance metric introduced in
Section~\ref{sec:formalism}.
 
\paragraph{Main result: the compositional bound.}
Let $G = (V, E, w)$ be an ICS detection network, $\pi = (f_{i_1}, \ldots, f_{i_k})$ a directed attack path in $G$, $\dist$ a disturbance class shared between the nodes, and $\tau \in (0,1]$ a performance threshold. The \emph{system-level absorption capacity} along $\pi$ is
$\mathcal{A}_{\mathrm{sys}}(\pi, \tau) = \sup\{b \geq 0 :
\varrho_{\mathrm{sys}}(\pi, b) \geq \tau\}$. We establish that
\begin{equation}
    \mathcal{A}_{\mathrm{sys}}(\pi,\, \tau)
    \;\leq\;
    \min_{j=1,\ldots,k}
    \frac{\mathcal{A}(f_{i_j},\, \dist,\, \tau)}{W(\pi,\, j)},
    \label{eq:comp_bound}
\end{equation}
where $\mathcal{A}(f_{i_j}, \dist, \tau)$ is the node-level absorption capacity of Eq.~\eqref{eq:absorption}. The bound is tight under the formal worst-case model of Eq.~\eqref{eq:sys_resilience}, where it coincides with $\mathcal{A}_{\mathrm{sys}}(\pi, \tau)$ exactly. We refer to it as the \emph{compositional bound} in what follows.
 
\paragraph{Intuitive justification.}
A disturbance of budget $b$ injected at the entry node $f_{i_1}$ delivers, through the path's coupling structure, an effective budget of at most $b \cdot W(\pi, j)$ at any downstream node $f_{i_j}$ (Eqs.~\eqref{eq:coupling_weight}--\eqref{eq:cumulative_coupling}). Node $f_{i_j}$ is compromised when its effective budget exceeds its absorption capacity, i.e. when $b \cdot W(\pi, j) > \mathcal{A}(f_{i_j}, \dist, \tau)$. Because the system metric in Eq.~\eqref{eq:sys_resilience} is a minimum over the path, the system is compromised as soon as any one node is, so the smallest entry budget compromising the path is exactly $\min_j \mathcal{A}(f_{i_j}, \dist, \tau)/W(\pi, j)$, which establishes the bound. Equality holds within the formal model; the looseness in real deployments arises because the realised propagated disturbance $\iota_{ij}(\delta)$ has its direction constrained by $\iota_{ij}$ rather than freely optimised, so physical attacks achieve less than the formal worst case. A complete derivation, including measurability of $\boldsymbol{\delta}^*$ and continuity of $\Phi$ under sequential disturbance propagation, is given in Appendix~\ref{app:proof}.
 
\paragraph{Interpretation.}
The bound~\eqref{eq:comp_bound} furnishes a \emph{system-level
resilience certificate}: a verdict on whether the whole path $\pi$
withstands a budget-$b_0$ adversary, issued from node-level
quantities and the coupling structure alone. Under the formal model
of Eq.~\eqref{eq:sys_resilience} the bound coincides with
$\mathcal{A}_{\mathrm{sys}}(\pi, \tau)$, so a single comparison against
an operationally relevant reference budget $b_0 > 0$ decides this
verdict in either direction. When the bound falls below $b_0$, the
certificate is negative, the system is fragile at $b_0$ along $\pi$, and the binding node $j^*(\pi) = \operatorname{argmin}_j
\mathcal{A}(f_{i_j})/W(\pi, j)$ identifies where a hardening
intervention tightens the bound most. When the bound meets $b_0$, the
certificate is positive, i.e. the system withstands every budget-$b_0$
attack along $\pi$, and this holds even when individual nodes carry
$\mathcal{A}(f_i) < b_0$, provided protective coupling raises each
$\mathcal{A}(f_i)/W(\pi, i)$ to $b_0$
(Section~\ref{subsec:eval_implications}). We call the bound
\emph{compositional} because both verdicts are determined by the path
topology and the coupling structure $\{w_{ij}\}$, which node-level
evaluation, seeing only the per-node capacities
$\{\mathcal{A}(f_i)\}$, cannot derive. The practical use is
correspondingly two-sided: compute the bound for every path
$\pi \in \Pi(G)$, then use it both to direct hardening to the binding
node on paths that fail certification and to certify paths that contain
sub-$b_0$ nodes whose contributions are shielded by attenuation.
 
Two structural properties of the bound are worth noting.
First, for any path with $W(\pi, j) = 1$ for all $j$ (perfectly lossless couplings), the bound reduces to the minimum node-level absorption capacity, recovering the classical weakest-link result as a special case.
Second, because $W(\pi, j)$ is non-increasing in $j$, deeper nodes are shielded by cumulative upstream attenuation: each downstream node receives effective budget $b \cdot W(\pi, j)$, which is smaller than the entry budget, and its contribution to the bound is $\mathcal{A}(f_{i_j})/W(\pi, j)$, which can exceed $\mathcal{A}(f_{i_j})$ by an arbitrary factor as $W$ shrinks. A node with higher individual absorption capacity but milder attenuation can therefore impose a tighter system-level bound than a weaker node positioned deeper in the path. Section~\ref{ex:three_node} exhibits this directly: $f_2$ with $\mathcal{A}(f_2) = 0.50$ binds at ratio $0.667$ while $f_3$ with $\mathcal{A}(f_3) = 0.30$ contributes ratio $0.800$ through the smaller $W(\pi, 3) = 0.375$, so the weaker node is the protected one. This is the formal mechanism by which the binding node becomes topology-dependent, and the reason why node-level evaluation, which observes only $\{\mathcal{A}(f_i)\}$, is structurally blind to it.
 
The bound extends \emph{classical series-system reliability}, the branch of reliability theory that analyses systems which fail as soon as any one component fails, where system reliability is the product of the component reliabilities and the system is dominated by its weakest component~\citep{Rausand2003}. This is the same min-over-the-path structure that Eq.~\eqref{eq:comp_bound} reduces to when all couplings are lossless ($W(\pi, j) = 1$). Classical series models, however, operate on failure probabilities and mean times to failure -- properties of component wear and random fault processes that are independent of any attacker strategy. The present bound instead operates on worst-case adversarial budgets and absorption capacities, properties of the adversarial-ML threat model that have no counterpart in the reliability literature, and it weights each component by the cumulative coupling $W(\pi, j)$ accumulated along the attack path. The binding constraint $\mathcal{A}(f_{i_j})/W(\pi, j)$ is consequently topology-dependent in a way that failure-probability models are not: the same node can be binding on one path and non-binding on another, depending on the accumulated coupling. This path-indexed, coupling-weighted structure is what distinguishes the result from both classical reliability theory and existing adversarial-ML evaluation, neither of which analyses the system-level resilience of a heterogeneous detection network as a function of attack-path topology.
\subsection{Implications for Adversarial Evaluation}
\label{subsec:eval_implications}
 
The bound in Eq.~\eqref{eq:comp_bound} has a direct and operationally significant implication for how adversarial robustness is evaluated in ICS deployments.
 
\paragraph{A consequence for hardening: insufficiency of node-level evaluation.}
A direct consequence of the compositional bound is that the binding node, the one whose individual hardening most tightens the system-level bound, need not be the minimum-capacity node identified by node-level evaluation. Formally, with $j_{\min} = \operatorname{argmin}_j \mathcal{A}(f_{i_j}, \dist, \tau)$ denoting the node-level minimum and
\begin{equation}
    j^* \;=\; \operatorname*{argmin}_{j=1,\ldots,k}
    \frac{\mathcal{A}(f_{i_j},\, \dist,\, \tau)}{W(\pi,\, j)}
    \label{eq:binding_mismatch}
\end{equation}
denoting the binding node of the bound, we have $j^* \neq j_{\min}$ in general, and the binding-node assignment $\pi \mapsto j^*(\pi)$ is itself path-dependent: the same network can have different binding nodes along different paths. Node-level adversarial evaluation produces only the values $\{\mathcal{A}(f_i, \dist, \tau)\}$, and therefore cannot identify $j^*$ without additional knowledge of the coupling weights $\{w_{ij}\}$ and the path structure. Consequently, it cannot direct hardening effort to the node whose improvement most tightens the system-level bound of Eq.~\eqref{eq:comp_bound}. The worked example below makes this concrete: the forward and reverse traversals of a single three-node network produce different binding nodes, neither of which coincides with the node-level minimum on the forward path.

\subsection{Three-node path example}
\label{ex:three_node}
\begin{SCfigure}[0.70][t]
    \includegraphics[width=0.38\linewidth]%
        {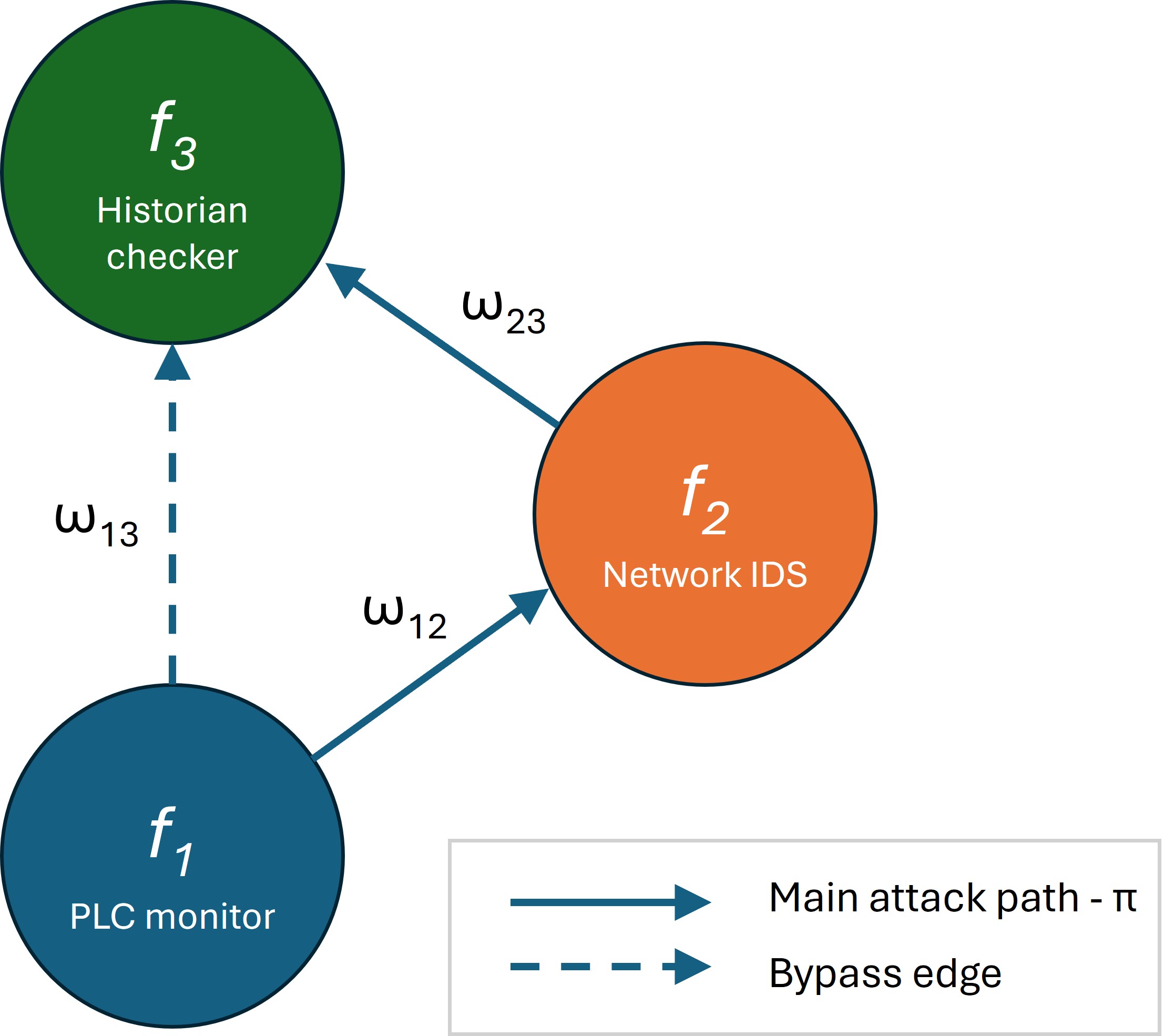}
    \caption{Example ICS detection network $G$ with three anomaly     detection nodes $f_1$ (PLC monitor), $f_2$ (network IDS), and $f_3$ (historian integrity checker). Directed edges carry coupling weights $w_{12}$, $w_{23}$, and $w_{13}$. The path $\pi = (f_1, f_2, f_3)$ has cumulative couplings
    $W(\pi,1) = 1$, $W(\pi,2) = w_{12}$,
    $W(\pi,3) = w_{12} w_{23}$.
    }
    \label{fig:ics_graph}
\end{SCfigure}

Consider an ICS detection network with three nodes representing a PLC process monitor ($f_1$), a network IDS ($f_2$), and a historian integrity checker ($f_3$), connected along a directed path $\pi = (f_1, f_2, f_3)$ (Figure~\ref{fig:ics_graph}).

Assign individual absorption capacities and coupling weights as follows:
\begin{equation}
    \mathcal{A}(f_1, \dist, \tau) = 0.85,\quad
    \mathcal{A}(f_2, \dist, \tau) = 0.50,\quad
    \mathcal{A}(f_3, \dist, \tau) = 0.30,
    \label{eq:ex_absorb}
\end{equation}
\begin{equation}
    w_{12} = 0.75,\quad w_{23} = 0.50,\quad \tau = 0.60.
    \label{eq:ex_weights}
\end{equation}
The values place $\mathcal{A}$ on the budget axis ($\ell_p$-norm units of the disturbance class) and $\tau$ on the performance axis (a value of $\Phi$). 
Node-level evaluation ranks
$f_3$ as the weakest ($\mathcal{A}(f_3) = 0.30$) and $f_1$ as the strongest. The cumulative coupling weights along $\pi$ are
\[
    W(\pi, 1) = 1.000,\quad
    W(\pi, 2) = 0.750,\quad
    W(\pi, 3) = 0.750 \times 0.500 = 0.375.
\]
Applying the bound~\eqref{eq:comp_bound}:
\begin{equation}
    \mathcal{A}_{\mathrm{sys}}(\pi, \tau)
    \;\leq\;
    \min\!\left(
    \frac{0.85}{1.000},\;
    \frac{0.50}{0.750},\;
    \frac{0.30}{0.375}
    \right)
    = \min(0.850,\; 0.667,\; 0.800)
    = 0.667.
    \label{eq:ex_bound1}
\end{equation}
The binding node is $f_2$ -- neither the node with the smallest absorption capacity ($f_3$, $\mathcal{A} = 0.30$) nor the one most exposed to the entry attack ($f_1$). This is the structural content of  Eq.~\eqref{eq:comp_bound}: the binding constraint is the ratio $\mathcal{A}/W$, not $\mathcal{A}$ alone. Despite $f_3$ being the weakest at the node level, the cumulative attenuation $W(\pi, 3) = 0.375$ insulates it from the entry threat: an attacker targeting $f_3$ must spend $0.30/0.375 = 0.800$ at the entry to deliver an effective budget of $0.30$ at $f_3$, and this exceeds the $0.667$ already sufficient to compromise $f_2$ through the milder $W(\pi, 2) = 0.75$ attenuation. Weak downstream coupling can therefore protect a fragile deep node, while moderate coupling at an intermediate position can expose a less-fragile mid-path node as the binding constraint.
 
\paragraph{Topology dependence.} The binding node is path-dependent. Consider the reverse direction $\pi' = (f_3, f_2, f_1)$ with coupling weights $w_{32} = 0.95$ and $w_{21} = 0.70$, reflecting tight historian-to-IDS and IDS-to-PLC alarm propagation:
\[
    W(\pi', 1) = 1.000,\quad
    W(\pi', 2) = 0.950,\quad
    W(\pi', 3) = 0.950 \times 0.700 = 0.665.
\]
\begin{equation}
    \mathcal{A}_{\mathrm{sys}}(\pi', \tau)
    \;\leq\;
    \min\!\left(
    \frac{0.30}{1.000},\;
    \frac{0.50}{0.950},\;
    \frac{0.85}{0.665}
    \right)
    = \min(0.300,\; 0.526,\; 1.278)
    = 0.300.
    \label{eq:ex_bound2}
\end{equation}
On the reverse path, $f_3$ is the entry node and is itself the binding constraint, because it now carries the full attack budget without prior attenuation. The same network therefore has binding node $f_2$ on $\pi$ and $f_3$ on $\pi'$: the binding-node assignment $\pi \mapsto j^*(\pi)$ is not a property of the nodes but of the path. The reverse path is also more vulnerable in absolute terms ($\mathcal{A}_{\mathrm{sys}} \leq 0.300$ versus $0.667$ on the forward path), entirely because moving the weakest node from a protected deep position to an exposed entry position removes its attenuation shield. Compositional analysis exposes both effects; node-level evaluation, which sees only $\{\mathcal{A}(f_i)\}$, registers neither.
\paragraph{Hardening implication.} On the forward path $\pi$, hardening $f_2$ is the only single-node intervention that tightens the bound of Eq.~\eqref{eq:comp_bound}. Hardening $f_1$ leaves the entry term $0.85/1.000$ above the binding ratio $0.667$ and so leaves $\min_j \mathcal{A}/W$ unchanged; hardening $f_3$ raises its ratio from $0.800$ but the bound is still set by $f_2$ at $0.667$. On the reverse path $\pi'$, by contrast, the corresponding intervention is hardening $f_3$ (the new entry/binding node). Routing hardening effort by node-level capacity -- "harden the weakest node first", namely $f_3$ -- is correct on the reverse path and actively wasted on the forward path, where the same capital improves no component of the system-level bound. 
 
\paragraph{Protective coupling and certification.}
The example also illustrates the converse direction of the node-vs-system relation: nodes that fail node-level evaluation can still contribute to a system that certifies at the same reference budget. Setting $b_0 = 0.60$, node-level evaluation rejects both $f_2$ ($\mathcal{A}(f_2) = 0.50 < 0.60$) and $f_3$ ($\mathcal{A}(f_3) = 0.30 < 0.60$), and would conclude that the network is fragile at $b_0$. The compositional bound gives $\mathcal{A}_{\mathrm{sys}}(\pi, 0.60) = 0.667 \geq 0.60$: the system is in fact resilient at $b_0$, because $f_2$ and $f_3$ are shielded by their cumulative couplings. An attacker spending the budget $b_0 = 0.60$ at the entry delivers only 
$0.60 \cdot W(\pi, j)$ at node $j$, which gives
$0.60 \cdot 0.750 = 0.45 < 0.50 = \mathcal{A}(f_2)$ and
$0.60 \cdot 0.375 = 0.225 < 0.30 = \mathcal{A}(f_3)$ --
both below the respective absorption thresholds. The example therefore exhibits both directions of node-vs-system mismatch: $f_3$ has the lowest node-level capacity but is not the binding node on $\pi$ (protective coupling), while $f_2$ has higher node-level capacity than $f_3$ but is the binding node (intermediate coupling exposes it). Both findings are properties of the path topology and are invisible to node-level evaluation.

\paragraph{Relationship to individual node analysis.}
Let $b_0 > 0$ denote a \emph{reference adversarial budget} on the budget axis, and say that a network \emph{passes node-level evaluation at $b_0$} when
\begin{equation}
    \mathcal{A}(f_i,\, \dist,\, \tau) \;\geq\; b_0
    \quad \text{for all } f_i \in V.
    \label{eq:all_robust}
\end{equation}
Under the formal model of Eq.~\eqref{eq:sys_resilience}, in which the bound of Eq.~\eqref{eq:comp_bound} is tight, three relations follow.
 
\emph{(i) Sufficient for certification.} Since $W(\pi, j) \in (0, 1]$ gives $\mathcal{A}/W \geq \mathcal{A}$, condition~\eqref{eq:all_robust} forces every term in the bound to lie at or above $b_0$; node-level pass implies system-level pass at the same $b_0$.
 
\emph{(ii) Not necessary for certification.} System-level pass at $b_0$ is equivalent to
\begin{equation}
    \mathcal{A}(f_{i_j},\, \dist,\, \tau)
    \;\geq\; b_0 \cdot W(\pi,\, j)
    \quad \text{for all } j,
    \label{eq:protective_coupling}
\end{equation}
strictly weaker than~\eqref{eq:all_robust} whenever any $W(\pi, j) < 1$. A node with $\mathcal{A}(f_{i_j}) < b_0$ can therefore contribute to a certifying system when cumulative coupling attenuates the attacker's reach below its absorption threshold.
 
\emph{(iii) Insufficient for hardening guidance.} The binding node $j^* = \operatorname{argmin}_j \mathcal{A}/W$ need not coincide with $j_{\min} = \operatorname{argmin}_j \mathcal{A}$ and is itself path-dependent
(Eq.~\eqref{eq:binding_mismatch}, Section~\ref{ex:three_node}); hardening targeted at $j_{\min}$ may leave $\mathcal{A}_{\mathrm{sys}}$ unchanged.
 
The compositional bound therefore both \emph{expands} the certification domain beyond node-level pass and \emph{redirects} hardening effort beyond the node-level minimum. The BATADAL deployment of Section~\ref{subsec:comp_validation} is a special case in which $j^* = j_{\min}$; Section~\ref{ex:three_node} shows this agreement is not generic. The standards-level consequences, i.e. coupling-aware certification clauses and hardening prioritised by $\mathcal{A}/W$ rather than $\mathcal{A}$, are developed in Section~\ref{subsec:certification}.

\section{Empirical Validation}
\label{sec:empirical}
 
The empirical validation draws on two complementary experimental foundations, inspired by our previous work~\citep{Stojanovic2022, Flatscher2024, Somma2024}: central-node anomaly detection and three-node (edge) anomaly detection in water distribution systems. 
The adversarial attack experiments on the three-node edge network, as well as extended central-node experiments, are \emph{new contributions of the present paper}, performed specifically to validate the compositional bound of Eq.~\eqref{eq:comp_bound}: these include white-box PGD evasion applied independently to the two-component central-node detector and to each edge detector, per-zone HoE computation (Section~\ref{subsec:comp_validation}), coupling weight estimation from BATADAL training data (Eq.~\eqref{eq:coupling_est}), and the adversarial training of one edge node ($f_2$) together with its compositional effect on the system-level bound. These experiments address the three validation objectives stated in Section~\ref{sec:intro}: formal metric estimation (Section~\ref{subsec:profiles}), recovery trajectory analysis (Section~\ref{subsec:recovery_empirical}), and compositional bound validation (Section~\ref{subsec:comp_validation}).
 
\subsection{Experimental Setup}
\label{subsec:setup}
 
\paragraph{Dataset.}
All experiments use the BATADAL dataset~\citep{Taormina2018}, a publicly available benchmark for cyber-attack detection in water distribution systems. The C-Town network comprises 429 pipes, 388 junctions, 7 storage tanks, 11 pumps across 5 stations, 5 valves, and 9 PLCs, instrumented with 43 SCADA channels sampled hourly. The test set (three months) contains 2\,089 samples: 1\,682 benign and 407 from 7 sustained attack campaigns of 30--100\,h each (Table~\ref{tab:campaign_hoe}). The campaign structure is exploited in Section~\ref{subsec:comp_validation} to derive campaign-level absorption capacity estimates under a sustained-attack threat model. 

\paragraph{Detection architectures.}
Two complementary architectures are evaluated.
 
\emph{(i) Single-node variants.}
Single-node architecture and experimental methodology are based on the adversarial robustness study of~\citet{Flatscher2024}, which studied adversarial robustness of the single-component detector based on AE for different training strategies under white-box PGD. In the given study~\citep{Flatscher2024} the base detector is an autoencoder: four dense layers with 43 neurons, 22-dimensional latent space, \texttt{tanh} activations, Adamax optimiser, early stopping (patience 10 epochs). 

In this paper, we extend the experiments of ~\citet{Flatscher2024}, by adding the second component to the detector, i.e. a smoothing filter: a post-hoc
7-sample moving average applied to the raw reconstruction error proposed in~\citep{Stojanovic2022}; this separation of the AE from the smoothing component is critical for the adversarial attack setup described below. Similarly to~\citep{Flatscher2024} the experiments in this paper include two central-node detector variants: $f_{\mathrm{orig}}$ -- trained using unmodified BATADAL data and $f_{\mathrm{adv}}$ -- trained with the adversarial loss.
Both variants are used in Section~\ref{subsec:profiles}.
 
\begin{figure}
    \centering
    \includegraphics[width=0.7\linewidth]{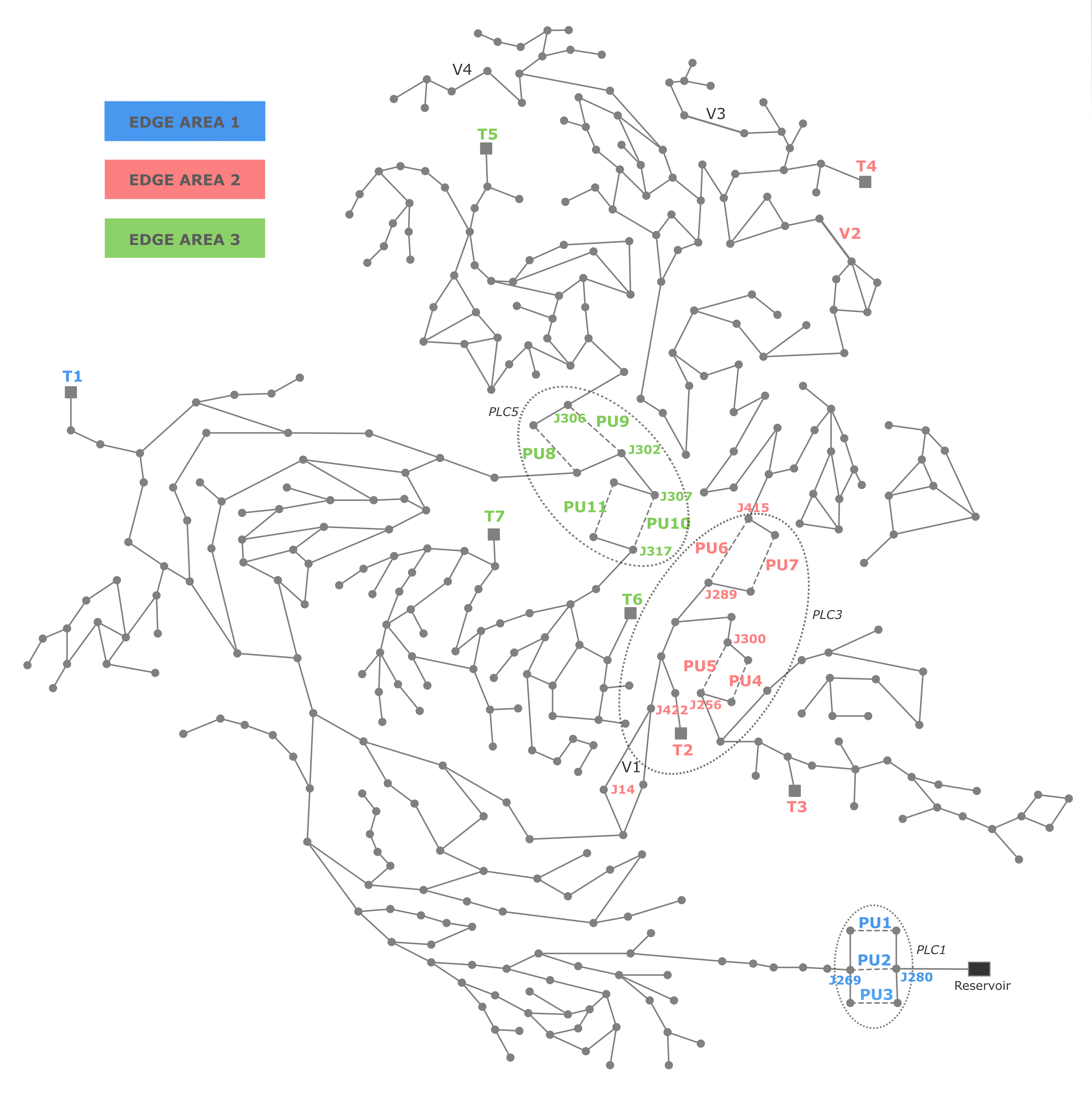}
    \caption{Network topology of the BADATAL water distribution system with partition in three edge areas and according PLC for edge device positioning.}
    \label{fig:batadal_wds_net}
\end{figure}

\emph{(ii) Three-node edge network.}
Three edge-node AEs trained independently on PLC-zone partitions of C-Town are based on the study of~\citet{Somma2024}. The same two-component detector architecture is adopted, with smoothing filter applied to reconstruction error for detection:
\begin{itemize}
    \item $f_1$ (Edge~1): 9 features, 5-dim latent;
          primary pumping station (L\_T1, PU1--PU3,
          P\_J280, P\_J269);
    \item $f_2$ (Edge~2): 19 features, 10-dim latent;
          central distribution (L\_T2--L\_T4, PU4--PU7,
          V2, six pressures);
    \item $f_3$ (Edge~3): 15 features, 8-dim latent;
          downstream zone (L\_T5--L\_T7, PU8--PU11,
          four pressures).
\end{itemize}
The BADATAL dataset is partitioned into three distinct areas based on the topology of the water distribution network. This division is directly aligned with the physical infrastructure, respecting the placement of PLCs at the network's edges. Fig.~\ref{fig:batadal_wds_net} illustrates the network topology with the defined edge areas. This careful segmentation mirrors the infrastructure of the water distribution system and the underlying physical processes, ensuring that the data division is both meaningful and practical.
Latent dimension is scaled to zone feature count~\citep{Somma2024}.

\paragraph{Attack scenarios.}
\emph{(i) Single-node variants (extended experiments).}
White-box PGD~\citep{Madry2018}, $\alpha = 10^{-4}$, $10\,000$ iterations. The attack objective is to minimise the 7-sample moving average of reconstruction error for each sample below the detection threshold, i.e.\ PGD finds the minimum-norm perturbation $\boldsymbol{\delta}$ such that
$\frac{1}{7}\sum_{s=t-6}^{t}\|f(x_s + \delta_s)\|^2 < \theta$.
Applied to all 407 attack-period samples across five feature-availability levels $k \in \{43, 33, 22, 11, 1\}$, where $k$ is the number of most-influential features available to the attacker~\citep{Flatscher2024}.
 
\emph{(ii) Edge-node variants (new experiments).}
The same PGD procedure applied independently to each edge AE at $k \in \{d_i,\lceil 0.75d_i\rceil , \lceil 0.5d_i\rceil ,\lceil  0.25d_i\rceil ,1\}$.
 
Table~\ref{tab:setup} summarises both configurations.
 
\begin{table}[t]
    \centering
    \caption{Experimental setup summary. Single-node
    variants for resilience profile estimation
    (Section~\ref{subsec:profiles}); edge-zone AEs for
    compositional validation
    (Section~\ref{subsec:comp_validation}).}
    \label{tab:setup}
    \renewcommand{\arraystretch}{1.2}
    \begin{tabular}{@{}lll@{}}
        \toprule
        \textbf{Parameter} &
        \textbf{Single-node} &
        \textbf{Edge-node} \\
        \midrule
        Dataset &
            BATADAL & BATADAL \\
        Architecture &
            AE, 4 layers, 22-dim latent &
            AE, 4 layers, 5/8/10-dim latent \\
        Variants &
            $f_{\mathrm{orig}}$, $f_{\mathrm{adv}}$ &
            $f_1$, $f_2$, $f_3$ (+$f_2^{\mathrm{adv}}$) \\
        Attack & White-box PGD & White-box PGD \\
        $\alpha$, iterations &
            $10^{-4}$, $10\,000$ &
            $10^{-4}$, $10\,000$ \\
        Feature levels &
            $k \in \{43,33,22,11,1\}$ &
            $k \in \{d_i,\lceil 0.75d_i\rceil ,
            \lceil 0.5d_i\rceil ,\lceil  0.25d_i\rceil ,1\}$ \\
        Primary metrics &
            HoE, $\hat{\Phi}$ &
            HoE, $\hat{\Phi}$, campaign-max \\
        References &
            \citep{Stojanovic2022,Flatscher2024}; Extended exp.: this paper &
            Arch.: \citep{Somma2024};
            Exp.: this paper \\
        \bottomrule
    \end{tabular}
\end{table}
 
\subsection{Measuring Resilience Profiles}
\label{subsec:profiles}
 
\paragraph{The anomaly detector as a two-component system.}
As previously stated, throughout this paper, the anomaly detector $f$ is defined as a \emph{complete system} consisting of two sequential components: the AE, which maps an input $x_t$ to a scalar reconstruction error $r_t = \|x_t - \hat{x}_t\|^2$; and a smoothing filter -- a moving-average filter of window $w = 7$, which smooths the raw reconstruction error into the anomaly score $\bar{r}_t = \frac{1}{7}\sum_{s=t-6}^{t} r_s$ that is compared against the detection threshold $\theta$. An alarm is raised when $\bar{r}_t > \theta$.
 
The study of~\citet{Flatscher2024} examined the adversarial robustness of the \emph{AE component alone}, without incorporating the averaging filter into the attack objective. The adversarial experiments reported in this paper target the \emph{full two-component detector} $f$: PGD minimises $\bar{r}_t$ (the smoothed score) rather than $r_t$ (the raw AE output), and the detection threshold $\theta$ is calibrated on $\bar{r}_t$. This distinction is operationally important: evasion of the AE component does not guarantee evasion of the full detector $f$ if the smoothed score remains above the threshold due to elevated errors in neighbouring time steps. Consequently, the HoE values and recall statistics reported below are properties of $f$ as a two-component system, not of the AE in isolation.
 
\paragraph{Mapping experimental metrics to resilience constructs.}
The new experiments yield two observables per detector variant per feature level: the Hardness of Evasion (HoE) and the successful attacks count. We map these onto the constructs of Section~\ref{sec:formalism} as follows.
 
\emph{Absorption capacity.}
By Eq.~\eqref{eq:absorption}, $\mathcal{A}(f, \mathcal{D}, \tau)$ is the supremum $\ell_2$ budget under which $f$ maintains recall at or above the operational performance threshold $\tau$. We instantiate this empirically as follows.
 
First, identify the critical feature level -- the supremum attacker capability at which the full detector still meets the performance criterion:
\begin{equation}
    k^*(\tau) \;=\; \sup\!\bigl\{k \;:\; \hat{\Phi}(f,\, k) \geq \tau\bigr\},
    \label{eq:k_star}
\end{equation}
where $\hat{\Phi}(f, k)$ is the empirical recall under minimum-norm PGD with $k$ features available to the attacker. At $k^*$, the detector is at the margin of acceptable performance: below $k^*$ the detector meets the criterion; above $k^*$ it does not. Formally, the feature-restricted disturbance class is the instantiation of Section~\ref{subsec:disturbance} with horizon $T = 1$, budget function $B(\delta) = \|\delta\|_2$, semantic constraint $\mathcal{C}(x) = \mathbb{R}^d$, and trajectory family $\mathcal{T}_k = \{\delta \in \mathbb{R}^d : \delta_i = 0\;\forall\,
i \notin S_k(f, x)\}$, where $S_k(f, x)$ is the index set of the $k$ input features with largest gradient magnitude $|\partial \hat{r}/\partial x_i|$ at the nominal input \citep{Flatscher2024}. Equivalently,
\begin{equation}
    \mathcal{D}_k(b) \;=\; \bigl\{ \delta \in \mathbb{R}^d :
    \|\delta\|_2 \leq b,\; \delta_i = 0\;\forall\, i \notin S_k(f, x)
    \bigr\},
    \label{eq:Dk_class}
\end{equation}
so each $\mathcal{D}_k$ is a proper instance of Section~\ref{subsec:disturbance} on which Eqs.~\eqref{eq:absorption} and~\eqref{eq:degradation} apply directly.

Second, estimate the absorption capacity as the HoE at this critical level:
\begin{equation}
    \hat{\mathcal{A}}(f,\, \tau)
    \;\approx\; \mathrm{HoE}\!\left(f,\, k^*(\tau)\right),
    \label{eq:absorb_estimate}
\end{equation}
where $\mathrm{HoE}(f, k)$ is defined as the median, over genuinely evaded samples $i$ in the attack-period set, of the per-sample \emph{normalised} (per-feature RMS) minimum $\ell_2$ evasion norm
\begin{equation}
    \eta_i(f, k) \;=\;
    \frac{1}{\sqrt{d}}
    \min \bigl\{ \|\delta\|_2 : \delta \in \mathcal{D}_k,\;
    \bar{r}_t(x_i + \delta) \leq \theta \bigr\},
    \label{eq:eta_def}
\end{equation}
where $d$ is the dimensionality of the detector's input feature space (so that $\eta_i$ has the operational interpretation of the RMS per-standardised-feature perturbation magnitude required to evade the smoothed detector), with $\bar{r}_t$ the smoothed
reconstruction error of the two-component detector, $\theta$ the detection threshold, and base false negatives ($\eta_i = 0$, i.e.\ samples already below $\theta$ without perturbation) excluded from the median.
The $\sqrt{d}$ normalisation makes $\hat{\mathcal{A}}$ comparable across detectors operating on feature spaces of different dimensionality (single-node $d = 43$ vs.\ edge zones $d_j \in \{9, 19, 15\}$).
The bridge to Eq.~\eqref{eq:absorption} is as follows: at $k^*$, evasion is possible at per-feature RMS cost $\mathrm{HoE}(f, k^*)$, so the detector can no longer maintain recall $\geq \tau$ once the attacker's normalised budget (per-feature RMS) exceeds this value. Hence $\mathrm{HoE}(f, k^*) \approx \mathcal{A}(f, \mathcal{D}_{k^*}, \tau)$ in the sense of Eq.~\eqref{eq:absorption}, with the disturbance class $\mathcal{D}_{k^*}$ instantiated as in Eq.~\eqref{eq:Dk_class} at the critical level and the budget axis interpreted in per-feature RMS units. 
 
\begin{figure}
    \centering
    \begin{subfigure}[t]{0.48\linewidth}
        \includegraphics[width=\linewidth]%
        {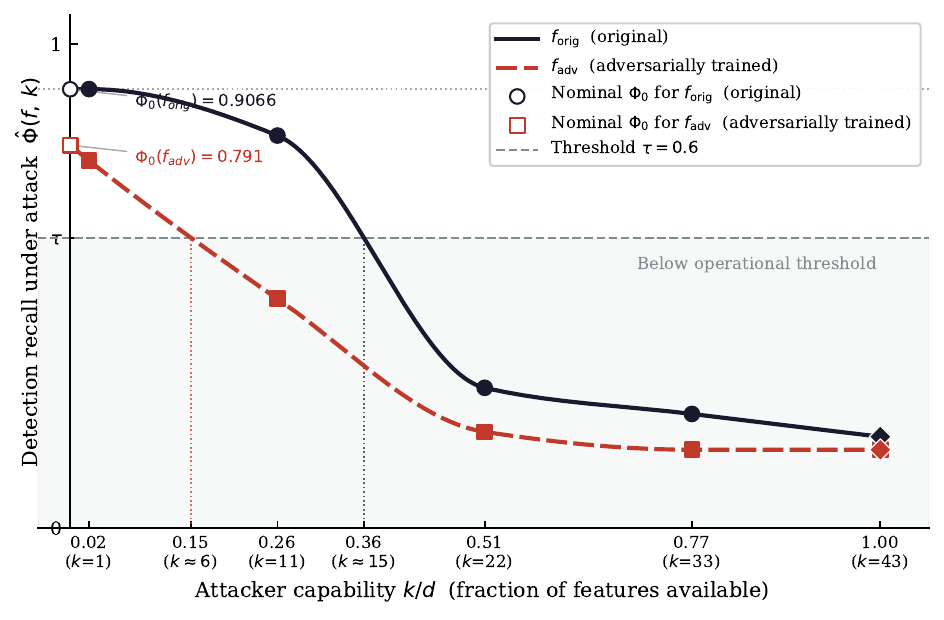}
        \caption{Empirical absorption curve:
        detection recall $\hat{\Phi}(f, k)$ under
        white-box PGD 
        vs.\ normalised
        attacker capability $k/d$.
        $f_{\mathrm{adv}}$ leads by per-sample HoE at every
        feature level 
        but lies below $f_{\mathrm{orig}}$
        in recall at every level, with critical feature level
        $k^*$ collapsing from $15$ to $6$ at $\tau = 0.60$. 
        }
        \label{fig:absorption_empirical}
    \end{subfigure}
    \hfill
    \begin{subfigure}[t]{0.48\linewidth}
        \includegraphics[width=\linewidth]%
        {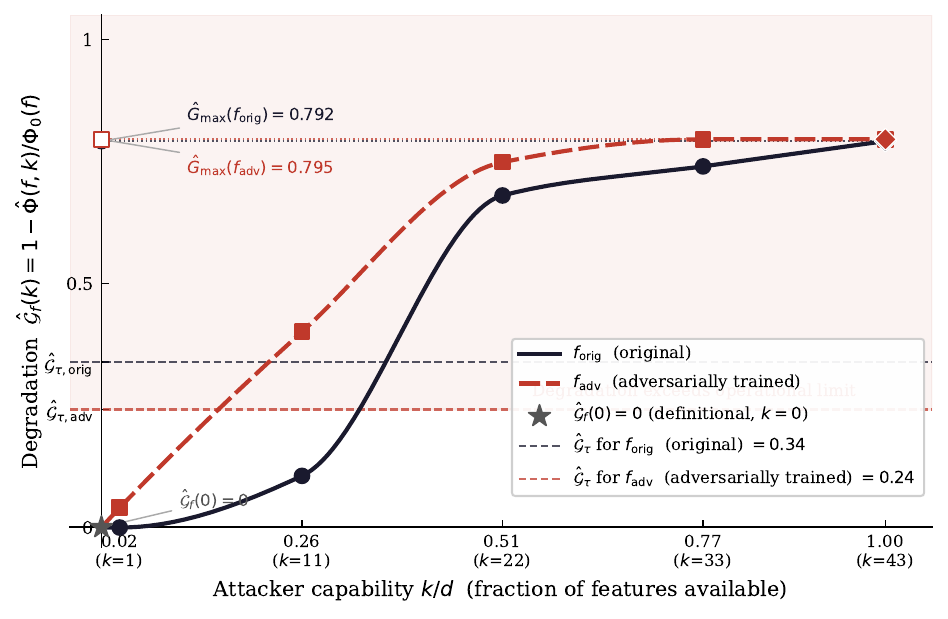}
        \caption{Empirical degradation function
        $\hat{\mathcal{G}}_f(k)$:
        each detector variant
        normalised against its own nominal recall.
        $f_{\mathrm{adv}}$'s degradation curve sits above
        $f_{\mathrm{orig}}$'s at every $k$ level.
        }
        \label{fig:degradation_empirical}
    \end{subfigure}
    \caption{Resilience profiles for the two single-node
    AE detector variants under white-box
    PGD on the BATADAL dataset. PGD targets the full
    two-component detector. 
    Neither panel alone provides a complete resilience
    characterisation: $f_{\mathrm{adv}}$ leads everywhere by
    HoE (Table~\ref{tab:resilience_profiles}) but lags
    everywhere by recall and degradation.}
    \label{fig:profiles}
\end{figure}

For operational threshold $\tau = 0.60$, the crossing point $k^*$ is read from the absorption curve; since measurements are at fixed $k$ levels, the true $k^*$ lies in the interval $[k_{\mathrm{lo}},\, k_{\mathrm{hi}})$ bounded by the two adjacent measured levels, and $\hat{\mathcal{A}}$ is estimated using PCHIP interpolation~\cite{Fritsch1984}, while HoE values are obtained experimentally (Figure~\ref{fig:profiles}).
 
\emph{Detection performance under attack.}
Recall under minimum-norm PGD targeting the full detector $f$:
\begin{equation}
    \hat{\Phi}(f, k) \;=\;
    1 - \frac{\text{successful evasions}(k)}{n},
    \label{eq:phi_estimate}
\end{equation}
where $n = 407$ is the total number of attack-period samples and ``successful evasion'' means $\bar{r}_t \leq \theta$ for the full two-component detector after perturbation. Two methodological notes apply. First, successful evasion counts conflate base false negatives (samples for which $\bar{r}_t \leq \theta$ without any perturbation) and genuine PGD evasions; base false negatives are identified from the unperturbed reconstruction errors and excluded from the HoE distribution. Second, the 407 attack samples span 7 sustained campaigns~\citep{Taormina2018}; samples within a campaign are
temporally autocorrelated, so the effective number of independent observations is approximately equal to the number of campaigns.
 
\emph{Degradation function.}
The nominal recall for the two centralised variants is measured
directly from the full-system PGD re-runs, with anomaly detection threshold calibrated to minimise the false positive rate: for $f_{\mathrm{orig}}$, $\Phi_0^{\mathrm{TPR}} = 0.9066$; for the adversarially trained
variant $f_{\mathrm{adv}}$, $\Phi_0^{\mathrm{TPR}} = 0.7912$.
The $f_{\mathrm{orig}}$ value supersedes the
$\Phi_0^{\mathrm{TPR}} = 0.9459$ reported by~\citet{Stojanovic2022},
which was measured on a detector threshold calibrated on a validation set to maximise the F1 score, balancing precision and recall.
Additionally, it should be noted that our experiments' re-runs do not extend the feature space with additional temporal cyclic features as proposed in ~\citet{Stojanovic2022}, because these features have no relevance for any perturbation method since they are a direct derivation of the timestamp and in a practical setting are calculated internally by the system, without possibility for interference. This approach ensures comparability between single-node and multi-node experiments, while a slight detection performance drop do not influence reasoning and results in any way.
The empirical degradation function is then
\begin{equation}
    \hat{\mathcal{G}}_f(k)
    \;=\; 1 - \frac{\hat{\Phi}(f,\, k)}
    {\Phi_0^{\mathrm{TPR}}(f)},
    \label{eq:degrad_estimate}
\end{equation}
computed with each detector's own $\Phi_0^{\mathrm{TPR}}(f)$ so that $\hat{\mathcal{G}}_f$ measures degradation relative to that detector's own nominal ceiling.
The estimator is parameterised by the attacker capability index $k$ rather than by an explicit $\ell_2$ budget $b$ for comparability between different models. 

Table~\ref{tab:resilience_profiles} collects all estimates for both $f_{\mathrm{orig}}$ and $f_{\mathrm{adv}}$ under the full-system PGD setup.
 
\begin{table}[t]
    \centering
    \caption{Empirical resilience profiles for the two central-node detector variants under the two-component full-system setup.
    $\hat{\mathcal{A}}(\tau{=}0.60) = \mathrm{HoE}(f, k^*)$:
    absorption capacity (Eqs.~\eqref{eq:k_star}--\eqref{eq:absorb_estimate})
    evaluated at the critical feature level $k^*(\tau{=}0.60)$. 
    $\dagger$~NaN: no genuine (non-zero) evasions at this level
    (all successful attacks are base false negatives).
    }
    \label{tab:resilience_profiles}
    \renewcommand{\arraystretch}{1.2}
    \begin{tabular}{@{}llccccc@{}}
        \toprule
        \textbf{Metric} & \textbf{Model}
        & \textbf{100\%} & \textbf{75\%}
        & \textbf{50\%} & \textbf{25\%} & \textbf{1feat} \\
        \midrule
        \multirow{2}{*}{\shortstack[l]{HoE\\
        (per-feat.\ RMS)}}
          & $f_{\mathrm{orig}}$
          & 0.1031 & 0.0801 & 0.0831 & 0.0345 & NaN$^\dagger$ \\
          & $f_{\mathrm{adv}}$
          & \textbf{0.1045} & \textbf{0.0979} & \textbf{0.0885}
          & \textbf{0.0731} & \textbf{0.0039} \\
 
        \midrule
        \multirow{2}{*}{\shortstack[l]{$\hat{\Phi}$\\
        (recall under attack)}}
          & $f_{\mathrm{orig}}$
          & \textbf{0.189} & \textbf{0.236} & \textbf{0.290}
          & \textbf{0.811} & \textbf{0.907} \\
          & $f_{\mathrm{adv}}$
          & 0.162 & 0.162 & 0.199 & 0.474 & 0.759 \\
 
        \midrule
        \multirow{2}{*}{\shortstack[l]{$\hat{\mathcal{G}}$\\
        (degradation)}}
          & $f_{\mathrm{orig}}$
          & \textbf{0.791} & \textbf{0.740} & \textbf{0.680}
          & \textbf{0.106} & \textbf{0.000} \\
          & $f_{\mathrm{adv}}$
          & 0.795 & 0.795 & 0.749 & 0.401 & 0.041 \\
        \midrule
        $\hat{\mathcal{A}}(\tau{=}0.60)$
          & $f_{\mathrm{orig}}$
          & \multicolumn{5}{c}{$\mathbf{0.0704}$\;\small{$k^*(0.60)\approx36\%(\approx15 feat.)$, $\hat{\Phi}(15)=\tau$}} \\
          & $f_{\mathrm{adv}}$
          & \multicolumn{5}{c}{$0.0471$\;\small{$k^*(0.60)\approx15\%(\approx 6 feat.)$, $\hat{\Phi}(6)=\tau$}}\\
        \bottomrule
    \end{tabular}
\end{table}
 
\paragraph{Key finding: absorption capacity and degradation are independent resilience characteristics.}
For $f_{\mathrm{orig}}$, the full two-component detector achieves HoE $= 0.1031$ at full features ($k=43$) with recall dropping to $\hat{\Phi} = 0.189$ ($\hat{\mathcal{G}} = 0.791$). The absorption capacity $\hat{\mathcal{A}}(f_{\mathrm{orig}}, \tau{=}0.60) = 0.0704$ (HoE at $k^*=15$, the critical feature level where $\hat{\Phi}$ first meets $\tau$).
At a single feature ($k=1$), PGD achieves zero genuine evasions, so $\hat{\Phi}(f_{\mathrm{orig}}, 1) = \hat{\Phi}(f_{\mathrm{orig}}, 0) = 0.907$ and $\hat{\mathcal{G}}(f_{\mathrm{orig}}, 1) = 0$. The moving-average component of $f$ provides effective single-feature resistance: a perturbation applied to only one sensor channel is smoothed across the 7-sample window and does not produce a sustained reduction in $\bar{r}_t$ large enough to evade the full detector.
 
The adversarially trained $f_{\mathrm{adv}}$ exhibits the accuracy--robustness trade-off characteristic of adversarial training~\citep{Tsipras2019, Madry2018} operating through the two-component detector, with the trade-off resolving against $f_{\mathrm{adv}}$ at the operational threshold. Per-sample median HoE rises modestly at every feature level (from $0.1031$ to $0.1045$ at $k=43$, and analogously through to $k=1$ where genuine evasions become possible at HoE $= 0.0039$, compared with no genuine evasions for $f_{\mathrm{orig}}$): adversarial training has succeeded in raising the per-sample evasion threshold of the AE component, as predicted by~\citet{Flatscher2024}. The composite behaviour goes the other way: the base false negative rate more than doubles, from $9.3\,\%$ for $f_{\mathrm{orig}}$ to $20.9\,\%$ for $f_{\mathrm{adv}}$, dropping nominal recall from $0.9066$ to $0.7912$ and depressing $\hat{\Phi}(f_{\mathrm{adv}}, k)$ below $f_{\mathrm{orig}}$'s recall at every feature level. The dominant effect is the base-rate degradation: at $\tau = 0.60$, the critical feature level for $f_{\mathrm{adv}}$ collapses from $k^* = 15$ to
$k^* = 6$, and the resulting absorption capacity falls  from $\hat{\mathcal{A}}(f_{\mathrm{orig}}, \tau{=}0.60) = 0.0704$ to $\hat{\mathcal{A}}(f_{\mathrm{adv}}, \tau{=}0.60) = 0.0471$.
This is the absorption--degradation divergence the resilience framework is designed to expose: the per-sample HoE distribution captures the geometry of the decision boundary (how much budget moves a nominally-detected sample below threshold), while $\hat{\mathcal{A}}$ at the operational threshold captures the composite operational behaviour (the budget at which the detector as a whole fails the operating criterion). Adversarial training improved the former and worsened the latter; a single-scalar robustness summary, in either direction, would mislead about the detector's operational standing.

Figure~\ref{fig:profiles} plots both resilience profiles (absorption and degradation) across the full feature-availability range.
It should be noted that the figures use $k/d$ as the x-axis because it parameterises the disturbance class $\dist_k$, i.e. the attacker's capability level. HoE is a scalar \emph{outcome} at each $(f, k)$ pair; using it as the x-axis would conflate the capability parameter with the
experimental result. 
 
\subsection{Recovery Trajectory}
\label{subsec:recovery_empirical}
 
\paragraph{The two-component anomaly detector produces a non-zero $T_r$.}
Both the single-node and edge architectures use an AE with a smoothing filter (7-sample moving average filter) applied to the raw reconstruction error, and the detection threshold $\theta$ is evaluated against this smoothed score. This design allows the PGD objective to target the smoothed score directly (see Section~\ref{subsec:setup}). When it comes to the recovery trajectory, it is important to note that when an attack ends at $t_0 + T$, the six preceding elevated-error samples remain in the moving-average window, influencing the smoothed score 
until $w - 1 = 6$ clean samples have been processed.
Formalising this (Eq.~\eqref{eq:recovery_traj}):
\begin{equation}
    T_r \;=\; w - 1 \;=\; 6 \text{ hours},\qquad
    \omega_r = 1.
    \label{eq:Tr_filter}
\end{equation}

Recovery time $T_r = 6$\,h (Eq.~\eqref{eq:recovery_time}) in our experiment applies only to detected attacks; for successfully evaded attacks the threshold was never crossed and recovery is instantaneous.
Recovery completeness value $\omega_r = 1$ (Eq.~\eqref{eq:recovery_completeness}) indicates full system recovery.
 
It should also be noted that $T_r$ is a structural property of the detector and it is \emph{orthogonal} to the divergent $\hat{\mathcal{A}}$ and $\hat{\mathcal{G}}$ profiles of Section~\ref{subsec:profiles}: detectors that differ radically on absorption and degradation are identical on recovery, confirming that $T_r$, $\mathcal{A}$, and $\mathcal{G}$ are independent resilience dimensions requiring separate measurement.

\paragraph{Filter size as a $T_r$--$\mathcal{A}$ trade-off.}
The filter-size experiment of~\citet{Stojanovic2022} can be reframed through the resilience lens. Each $w$ tested ($w \in \{1, 3, 5, 7, 9, 11 \}$) defines
a $(T_r, S_{\mathrm{CLF}})$ operating point: larger $w$ improves
balanced accuracy at the cost of longer recovery, while
$S_{\mathrm{TTD}}$ is invariant to $w$. The final selection $w = 7$ was set to maximise the F1 score.
 
\paragraph{Temporal detection performance: edge vs centralised.} \citet{Somma2024} investigates $S_{\mathrm{TTD}}$ (time from attack onset to first detection). The edge deployment achieves $S_{\mathrm{TTD}} = 0.9701$ vs.\ $0.9556$ for the centralised model ($+1.5\,\%$), while $S_{\mathrm{CLF}}$ is lower (0.9469 vs.\ 0.9587); both share $T_r = 6$\,h (Table~\ref{tab:recovery_comparison}).
 
$S_{\mathrm{TTD}}$, $S_{\mathrm{CLF}}$, and $T_r$ are three distinct scalar summaries of $\recov(f, t_0, \dist)$, each capturing a different operational characteristic and confirming that the three resilience dimensions are independent.
 
\begin{table}[t]
    \centering
    \caption{Three independent resilience dimensions.
    $S_{\mathrm{TTD}}$ maps to detection latency;
    $S_{\mathrm{CLF}}/F_1/$TPR map to $\mathcal{A}$;
    $T_r = 6$\,h 
    maps to recovery time.}
    \label{tab:recovery_comparison}
    \renewcommand{\arraystretch}{1.2}
    \begin{tabular}{@{}lccccc@{}}
        \toprule
        \textbf{Architecture} &
        $S_{\mathrm{TTD}}$ & $S_{\mathrm{CLF}}$ &
        $F_1$ & TPR & $T_r$ \\
        \midrule
        Centralised AE~\citep{Stojanovic2022}
          & 0.9556 & 0.9587 & 0.9167 & 0.9459 & 6\,h \\
        Edge AE~\citep{Somma2024}
          & \textbf{0.9701} & 0.9469
          & 0.9253 & 0.9056 & 6\,h \\
        \midrule
        \textit{Maps to:}
          & Latency & $\mathcal{A}$
          & $\mathcal{A}$ & $\mathcal{A}$ & $T_r$ \\
        \bottomrule
    \end{tabular}
\end{table}
 
\subsection{Compositional Validation}
\label{subsec:comp_validation}
 
We construct a three-node ICS detection network from the edge-zone deployment of~\citet{Somma2024} and conduct new adversarial evasion experiments to verify the compositional bound of Eq.~\eqref{eq:comp_bound}. The edge AE architectures 
follow~\citet{Somma2024}; the PGD evasion experiments, HoE computation, coupling weight estimation, and adversarial training of $f_2$ are new contributions of this paper.
 
\paragraph{Network structure and coupling weights.}
We model the three edge zones as nodes in $G = (\{f_1, f_2, f_3\}, E, w)$ following the hydraulic flow structure of C-Town~\citep{Taormina2018}: primary pumping station (Edge~1) $\to$ central distribution (Edge~2) $\to$ downstream zone (Edge~3), giving directed edges $(f_1, f_2)$ and $(f_2, f_3)$.
 
Coupling weights are estimated from raw SCADA sensor readings of the normal-operation training set. 
Binary status columns ($S\_PU_k$, $S\_V_k$) are excluded. The coupling weight $\hat{w}_{ij}$ is the maximum absolute forward-lag cross-correlation over physically motivated boundary sensor
pairs $\mathcal{S}_{ij}$:
\begin{equation}
    \hat{w}_{ij} = \max_{(c_i, c_j) \in \mathcal{S}_{ij}}
    \max_{\ell \in \{0,\ldots,3\}}
    |\mathrm{corr}(c_{i,t},\; c_{j,\,t+\ell})|.
    \label{eq:coupling_est}
\end{equation}
Boundary pairs are selected from the C-Town topology~\citep{Taormina2018}
and PLC-zone partition~\citep{Somma2024}
based on the hydraulic adjacency criterion.
For edge $(f_1, f_2)$: (F\_PU3, F\_PU4), (F\_PU3, F\_PU6),
(P\_J280, P\_J300), (P\_J269, P\_J256), (L\_T1, L\_T2),
(L\_T1, L\_T3).
For edge $(f_2, f_3)$: (L\_T3, L\_T5), (L\_T4, L\_T5),
(L\_T4, L\_T6), (L\_T3, L\_T6), (F\_PU7, F\_PU8),
(P\_J415, P\_J302).
From the BATADAL training data:
\begin{equation}
    \hat{w}_{12} = 0.422,\qquad \hat{w}_{23} = 0.319,
    \label{eq:coupling_values}
\end{equation}
yielding cumulative coupling:
\begin{equation}
    W(\pi,1) = 1.000,\quad
    W(\pi,2) = 0.422,\quad
    W(\pi,3) = 0.422 \times 0.319 = 0.135.
    \label{eq:cumulative_coupling_values}
\end{equation}
A disturbance must carry approximately $7.4\times$ the entry budget
to induce an equivalent perturbation at $f_3$. Coupling estimates
vary by less than $0.03$ across lags $1\ldots3$\,h.
 
\paragraph{Empirical estimator vs formal Lipschitz definition.}
The formal coupling weight $w_{ij}$ in Eq.~\eqref{eq:coupling_weight}
is a worst-case adversarial quantity: a Lipschitz-style supremum of
the propagation map $\iota_{ij}$ over all admissible disturbances.
The empirical estimator $\hat{w}_{ij}$ in Eq.~\eqref{eq:coupling_est}
is the maximum absolute Pearson
cross-correlation of boundary sensor channels under \emph{normal}
operation. Disturbance propagation in the SCADA-controlled deployment
operates through two mechanisms: (i)~\emph{control-loop mediated
coupling}, in which a spoofed reading triggers an actuator command
that physically alters the downstream process; and (ii)~\emph{shared
sensor coupling}, in which a sensor reading appearing in both feature
sets is simultaneously corrupted, contributing a term of $1$ to the
supremum in Eq.~\eqref{eq:coupling_weight}. The cross-correlation
estimator is a heuristic surrogate that approximates the linear
component of the control-loop channel under the standing assumption
that adversarial signal propagation does not exceed the dynamic
range observed in normal operation.
Quantifying the gap between the Pearson estimate and the true
Lipschitz constant of $\iota_{ij}$ under nonlinear hydraulic dynamics
is identified as an open problem.
 
\paragraph{Cross-zone comparability of HoE under normalisation.}
Throughout this section HoE values are reported in per-feature RMS
units (Eq.~\eqref{eq:eta_def}: $\eta_i = \|\delta\|_2/\sqrt{d_j}$),
not in raw $\ell_2$ norm. Under standardised feature preprocessing, a
uniform per-feature perturbation of magnitude $\varepsilon$ produces
$\|\delta\|_2 = \varepsilon\sqrt{d_j}$ in raw units but $\eta_i =
\varepsilon$ under the normalisation, so per-feature RMS HoE values
are directly comparable across the three edge zones ($d_1{=}9$,
$d_2{=}19$, $d_3{=}15$) and against the centralised AE ($d{=}43$)
of Section~\ref{subsec:profiles}. The normalisation is applied
per-zone in the path bound: each term
$\hat{\mathcal{A}}(f_j, \tau)/W(\pi, j)$ in
Eq.~\eqref{eq:comp_bound} uses $f_j$'s own
$\sqrt{d_j}$ factor, so the budget axis is consistently the
per-feature RMS axis along the path. 
 
\paragraph{Campaign-maximum HoE and the sustained-attack threat model.}
HoE values are reported in per-feature RMS units
(per-feature RMS units; see the cross-zone-comparability note above).
The formal absorption capacity $\hat{\mathcal{A}}(f_j, \tau)$
(Eqs.~\eqref{eq:k_star}--\eqref{eq:absorb_estimate})
is derived from the per-sample median HoE at $k^*$.
A complementary, operationally relevant quantity for a
sustained-attack threat model is the \emph{campaign-maximum} HoE
-- the minimum evasion norm of the hardest sample in the
campaign window, i.e.\ the budget an attacker must sustain to
evade every sample:
\begin{equation}
    \hat{\mathcal{A}}_c(f_j)
    \;=\; \max_{i \in \mathcal{I}_c} \eta_i(f_j),
    \label{eq:campaign_max}
\end{equation}
where $\eta_i(f_j)$ is the per-feature RMS minimum evasion norm
for sample $i$ (Eq.~\eqref{eq:eta_def}; zero for base false
negatives, excluded from the maximum).
The campaign-level absorption capacity is the median of
$\hat{\mathcal{A}}_c$ over campaigns targeting zone $j$. An
adversary must evade every sample in a campaign window, not only
the median one.
 
Analysis of campaign timestamps confirms that each BATADAL attack
campaign targets sensors in exactly one zone
(Table~\ref{tab:campaign_hoe}), independently validating the PLC-zone
partition. 
 
\begin{table}[t]
    \centering
    \caption{Campaign-level HoE analysis for each BATADAL test
    attack~\citep{Taormina2018}. Campaign-max HoE 
    is the budget required to evade all
    samples in the campaign window. Dashes (---): attack outside zone's feature set.
}
    \label{tab:campaign_hoe}
    \renewcommand{\arraystretch}{1.2}
    \begin{tabular}{@{}clccccc@{}}
        \toprule
        \textbf{C} & \textbf{Target} & \textbf{$n$} &
        \multicolumn{2}{c}{$f_2$ (Edge~2)} &
        \multicolumn{2}{c}{$f_3$ (Edge~3)} \\
        \cmidrule(lr){4-5}\cmidrule(lr){6-7}
        & & & evaded & camp.-max & evaded & camp.-max \\
        \midrule
        8  & Edge~2 & 70  & 50/70  & 0.493 & 0/70  & ---   \\
        9  & Edge~2 & 65  & 54/65  & 0.590 & 0/65  & ---   \\
        10 & Edge~1 & 31  & 0/31 & ---  & 0/31  & ---   \\
        11 & Edge~1 & 31  & 0/31 & ---  & 0/31  & ---   \\
        12 & Edge~2 & 100 & 91/100 & 1.336 & 0/100 & ---   \\
        13 & Edge~3 & 80  & 0/80   & ---   & 76/80 & 0.451 \\
        14 & Edge~2 & 30  & 8/30   & 3.143 & 0/30  & ---   \\
        \midrule
        \multicolumn{3}{l}{Campaign-median (targeted)} &
        \multicolumn{2}{c}{$0.980$} &
        \multicolumn{2}{c}{$0.451$} \\
        \bottomrule
    \end{tabular}
\end{table}
 
\paragraph{Edge~1: complete structural separation.}
Table~\ref{tab:campaign_hoe} and Table~\ref{tab:comp_absorb} show that PGD fails to evade detection on \emph{any} of the 62 attack-period
samples at any feature level: $\hat{\Phi}(f_1, k) = 1.000$ for all
$k \in \{d_1, \lceil 0.75 d_1\rceil, \ldots, 1\}$. The nominal
recall $\hat{\Phi}(f_1, 0) = 1.000$ confirms zero base false negatives
($0/62$ samples below threshold without perturbation). The post-PGD
recall is identical to the nominal recall at every feature level,
confirming that PGD has zero effect on Edge~1 regardless of the
perturbation budget. Campaigns C10 and C11 (PU3 activation) produce
zero evasion norms at both $f_2$ and $f_3$. The NaN result formally
confirms $\hat{\mathcal{A}}(f_1) \to \infty$: $f_1$ is non-binding
in Eq.~\eqref{eq:comp_bound}.
 
\paragraph{Structural vs adversarial robustness.}
Edge~1's non-evadability is not general robustness. The resilience
framework separates $\Phi_0$ (nominal detection capability) from
$\mathcal{A}$ (resistance to deliberate evasion); here both are maximal
($\Phi_0 = 1.000$, $\mathcal{A} \to \infty$), making $f_1$ the
structurally strongest node in the network despite operating on the
smallest feature set ($d_1 = 9$).
 
\paragraph{Edge~2 and Edge~3: vulnerability and detection profiles.}
The nominal recall of Edge~2, $\hat{\Phi}(f_2, 0) = 0.925$, indicates
that $7.5\,\%$ of zone-specific attack samples fall below the
detection threshold without adversarial perturbation. Under full-feature
PGD, recall drops from $0.925$ to $0.117$.
Recall recovers substantially as
attacker capability decreases: $0.536$ at $50\,\%$ features and
$0.909$ at a single feature ($k=1$).
 
Edge~3 achieves complete evasion at full features
($\hat{\Phi}(f_3, 0) = 1.000$, $\hat{\Phi}(f_3, d_3) = 0.000$
.
At $50\,\%$ features recall recovers to $0.425$ and reaches $\hat{\Phi}(f_3, 1) = 1.000$ at a single feature
-- complete recovery of the nominal recall. 
Table~\ref{tab:comp_absorb} consolidates the per-sample statistics,
including results for the adversarially trained $f_2^{\mathrm{adv}}$.
 
\begin{table}[t]
    \centering
    \caption{Per-zone PGD summary. All HoE values
    are per-feature RMS norms (Eq.~\eqref{eq:eta_def};
    per-feature RMS units; see the cross-zone-comparability note above).
    $\hat{\Phi}(k{=}0)$ is nominal recall (no perturbation).
    Camp.-max is campaign-maximum HoE 
    (Eq.~\eqref{eq:campaign_max}).
    $\hat{\mathcal{A}}(\tau{=}0.60)=\mathrm{HoE}(k^*)$ is absorption
    capacity (Eqs.~\eqref{eq:k_star}--\eqref{eq:absorb_estimate});
    $k^*$ identified per zone from the absorption curve at $\tau=0.60$.
    $\dagger$~NaN/\!$\infty$: no genuine evasions ($f_1$ absorbs all
    attacks). $\ddagger$~zero: nominal recall below $\tau$ at $k{=}1$.}
    \label{tab:comp_absorb}
    \renewcommand{\arraystretch}{1.2}
    \begin{tabular}{@{}lccccccccc@{}}
        \toprule
        \textbf{Node} & $d_i$ &
        \multicolumn{4}{c}{$\hat{\Phi}$ (recall)} &
        \multicolumn{2}{c}{HoE (per-feat.\ RMS)} &
        Camp.-max &
        $\hat{\mathcal{A}}(\tau{=}0.60)$ \\
        \cmidrule(lr){3-6}\cmidrule(lr){7-8}
        & &
        $k{=}0$ & $k{=}d_i$ &
        $k{=}\lceil 0.5d_i\rceil$ & $k{=}1$ &
        $k{=}d_i$ & $k{=}\lceil 0.5d_i\rceil$ &
        med. & HoE$(k^*)$ \\
        \midrule
        $f_1$ (Edge~1) & 9
          & 1.000 & 1.000 & 1.000 & 1.000
          & NaN$^\dagger$ & NaN$^\dagger$ & $\infty^\dagger$ \\
        $f_2$ (Edge~2) & 19
          & 0.925 & 0.117 & 0.536 & 0.909
          & 0.123 & 0.003 & 0.980 & 0.088 \\
        $f_2^{\mathrm{adv}}$ & 19
          & 0.687 & 0.023 & 0.125 & 0.585
          & 0.218 & 0.197 & 0.842 & $0^\ddagger$ \\
        $f_3$ (Edge~3) & 15
          & 1.000 & 0.000 & 0.425 & 1.000
          & 0.183 & 0.096 & 0.451 & 0.133 \\
        \bottomrule
    \end{tabular}
\end{table}
 
\paragraph{Adversarially trained Edge~2: binding node confirmed,
divergence replicated.}
Adversarial training of $f_2$ (identical to $f_{\mathrm{adv}}$ in Section~\ref{subsec:profiles})
increases the base false negative rate from $7.5\,\%$ to
$31.3\,\%$: the adversarially trained model nominally misses
nearly a third of zone-specific attack samples before any perturbation
is applied, and the nominal recall drops from $0.925$ to $0.687$.
The per-sample median HoE increases from $0.123$ to $0.218$ at full features (per-feature RMS), consistent with
the adversarial training raising the evasion threshold, but this gain
is more than offset by the $23.8\,\%$ increase in base
false negative rate.
 
This result replicates the absorption--degradation divergence of
Section~\ref{subsec:profiles} at the edge-zone level, and is
consistent with the provably fundamental accuracy--robustness
trade-off~\citep{Tsipras2019, Madry2018}: adversarial training raises
the per-sample evasion threshold while simultaneously degrading nominal
detection capability, without shifting the binding constraint.

\paragraph{Compositional bound.}
Table~\ref{tab:comp_bound_result} applies Eq.~\eqref{eq:comp_bound}
using the absorption capacity $\hat{\mathcal{A}}(f_j, \tau{=}0.60)
= \mathrm{HoE}(f_j, k^*)$ (Eqs.~\eqref{eq:k_star}--\eqref{eq:absorb_estimate})
with $k^*$ identified from the discrete absorption curve for each zone.

\paragraph{Boundary case $k^* = 0$.}
The reading $\hat{\mathcal{A}}(f_2^{\mathrm{adv}}) = 0$
in Table~\ref{tab:comp_bound_result} arises because the supremum in
Eq.~\eqref{eq:k_star} is taken over the discrete grid
$k \in \{1, \lceil 0.25 d_j \rceil, \ldots, d_j\}$: for
$f_2^{\mathrm{adv}}$, the smallest non-zero capability level $k = 1$
already drives recall below $\tau$, so $\sup\{k : \hat{\Phi}(f, k)
\geq \tau\} = 0$. The continuous-$b$ formal capacity
$\mathcal{A}(f_2^{\mathrm{adv}}, \mathcal{D}_1, \tau)$ in
Eq.~\eqref{eq:absorption} is also zero in this case, since
arbitrarily small $\ell_2$ budget under $\mathcal{D}_1$ already
suffices to cross $\tau$. The empirical and formal values therefore
agree at this boundary; the $0$ entry should be read as a
\emph{floor} on the binding ratio rather than an artefact of the
discrete grid.

\begin{table}[t]
    \centering
    \caption{Compositional bound for path $\pi=(f_1,f_2,f_3)$
    using absorption capacity $\hat{\mathcal{A}}(\tau{=}0.60)
    =\mathrm{HoE}(f_j, k^*)$ in per-feature RMS units
    (Eqs.~\eqref{eq:k_star}--\eqref{eq:absorb_estimate};
    per-feature RMS units; see the cross-zone-comparability note above).
    $f_1$ absorbs all feature levels
    ($\hat{\mathcal{A}}\to\infty$, non-binding). $f_2$ is the binding node.
    Adversarially trained $f_2^{\mathrm{adv}}$ has nominal recall
    $0.687 \geq \tau=0.60$ (nominally acceptable), yet
    even a single-feature attack yields $\hat{\Phi}=0.585 < \tau$,
    so $k^*=0$ and $\hat{\mathcal{A}}(f_2^{\mathrm{adv}},0.60)=0$:
    the model absorbs zero adversarial capability.
    }
    \label{tab:comp_bound_result}
    \renewcommand{\arraystretch}{1.2}
    \begin{tabular}{@{}lcccc@{}}
        \toprule
        \textbf{Node} &
        $\hat{\mathcal{A}}(\tau{=}0.60)$\;[HoE$(k^*)$] &
        $W(\pi,j)$ &
        $\hat{\mathcal{A}} / W(\pi,j)$ &
        \textbf{Binding?} \\
        \midrule
        $f_1$ (Edge~1)
          & $\infty$ (NaN) & 1.000 & $\infty$ & No \\
        \midrule
        $f_2$ (Edge~2)
          & $0.0883$\;\small{$k^*=4$} & 0.422
          & \textbf{0.209} & \textbf{Yes} \\
        \midrule
        $f_2^{\mathrm{adv}}$
          & $0$\;\small{nom.~$\hat{\Phi}_0\geq\tau$, but $k^*=0$} & 0.422
          & $0$ & Yes \\
        \midrule
        $f_3$ (Edge~3)
          & $0.1330$\;\small{$k^*=4$} & 0.135
          & 0.988 & No \\
        \midrule
        \multicolumn{2}{l}{$\mathcal{A}_{\mathrm{sys}}(\pi,\,\tau{=}0.60)$\;($f_2$)\,$\leq$}
          & & \textbf{0.209} & \\
        \bottomrule
    \end{tabular}
\end{table}
 
Table~\ref{tab:comp_bound_result} confirms that $f_2$ is the binding
node: $\hat{\mathcal{A}}(f_2)/W(\pi,2) = 0.0883/0.422 = 0.209$, while
$f_3$'s term ($0.1330/0.135 = 0.988$) is substantially higher despite
$f_3$ being fully evadable at full features. The binding constraint is
therefore determined by the ratio $\hat{\mathcal{A}}/W(\pi,j)$,
not by $\hat{\mathcal{A}}$ alone, a direct empirical confirmation
of the compositional bound structure of
Eq.~\eqref{eq:comp_bound}. All quantities are expressed in
per-feature RMS units (per-feature RMS units; see the cross-zone-comparability note above); the binding-node
identification is preserved under per-zone normalisation, and the
per-feature interpretation makes
$\mathcal{A}_{\mathrm{sys}}(\pi, \tau{=}0.60) \leq 0.209$
directly comparable across the three edge zones and against the
centralised AE result
$\hat{\mathcal{A}}(f_{\mathrm{orig}}, \tau{=}0.60) = 0.0704$ of
Section~\ref{subsec:profiles}.
 
The edge deployment's $100\,\%$ attack localisation
accuracy~\citep{Somma2024} provides independent qualitative support:
attacks targeting Edge~2 sensors leave Edge~3 naturally unaffected,
and vice versa, validating the PLC-zone partition as a hydraulically
meaningful security boundary.

\subsection{Architectural Comparison: Centralised vs.\ Three-Zone Edge}
\label{subsec:arch_comparison}
 
The resilience constructs of Section~\ref{sec:formalism} provide a
common set of axes along which the two architectural baselines can
be compared. The per-feature RMS normalisation
(per-feature RMS units; see the cross-zone-comparability note above) makes absorption capacities directly
comparable across the centralised feature space ($d = 43$) and the
partitioned edge feature spaces ($d_j \in \{9, 19, 15\}$); each of
the four framework constructs then yields a specific architectural
comparison.
 
\paragraph{Absorption capacity at the operational threshold.}
At $\tau = 0.60$, the centralised AE achieves
$\hat{\mathcal{A}}(f_{\mathrm{orig}}, \tau) = 0.0704$ (per-feature
RMS, Table~\ref{tab:resilience_profiles}); the three-zone edge
system achieves $\mathcal{A}_{\mathrm{sys}}(\pi, \tau) = 0.209$ via
the compositional bound (Table~\ref{tab:comp_bound_result}), set by
the binding node $f_2$ through $\hat{\mathcal{A}}(f_2)/W(\pi, 2)
= 0.0883/0.422$. The edge architecture therefore admits a
$\sim 3\times$ larger per-feature adversarial budget at the same
operational threshold before recall falls below $\tau$. Two
structural mechanisms account for the gap. Per-zone feature
partitioning narrows the dimensionality available to a single-zone
attacker, raising the per-feature attack cost in the binding zone
$f_2$ ($d_2 = 19$) relative to the centralised ($d = 43$).
Cumulative coupling attenuation $W(\pi, j) < 1$ then shields
downstream zones from any budget injected at the entry: $f_3$
contributes ratio $0.988$ to the bound through its small $W(\pi, 3)
= 0.135$, despite a node-level $\hat{\mathcal{A}}(f_3) = 0.133$ that
is itself larger than the centralised capacity. The non-binding
ratio $0.988$ is roughly $4.7\times$ larger than the binding ratio
$0.209$, which means $f_2$ is the sole driver of the system-level
constraint and the remaining headroom at $f_3$ is unrealised in
the bound.
 
\paragraph{Degradation profile and adversarial training.}
Both architectures exhibit the absorption-degradation divergence
under adversarial training, with structurally different system-level
consequences. For the centralised AE the divergence is internal to
a single detector: $\hat{\mathcal{A}}(f_{\mathrm{adv}}, \tau) =
0.0471$ is a reduction relative to
$f_{\mathrm{orig}}$ (Section~\ref{subsec:profiles}), and one of the possible remediation could be detector replacement. For the edge
architecture, adversarial training of the binding node $f_2$
collapses the system bound to
$\mathcal{A}_{\mathrm{sys}}(\pi, \tau) = 0$
(Section~\ref{subsec:comp_validation}) because the loss of $f_2$'s
absorption capacity propagates linearly through $W(\pi, 2)$. The
collapse is more severe at the system level than the 
reduction observed centrally, but the architecture localises the
failure mode: $f_1$ and $f_3$ are unaffected, the original bound
is recoverable through a change confined to $f_2$, and the
compositional analysis identifies $f_2$ as the precise point
of remediation. Compositional certification therefore not only
quantifies the system-level effect of a per-node intervention
but identifies where in the network remediation needs to act.
 
\paragraph{Recovery trajectory and detection latency.}
The two architectures share the same recovery horizon
$T_r = 6\,\mathrm{h}$ and recovery completeness $\omega_r = 1$
(Section~\ref{subsec:recovery_empirical}), both fixed by the
post-hoc smoothing filter applied uniformly in both deployments.
Architectural distribution does not change $T_r$ but improves
detection latency: $S_{\mathrm{TTD}} = 0.9701$ for the edge
architecture against $0.9556$ for the centralised
(Table~\ref{tab:recovery_comparison}), a $1.45$\% absolute
improvement that is independent of $T_r$ and arises from
per-zone specialisation of the reconstruction objective.
 
\paragraph{Disturbance class scope.}
The disturbance class at the input level coincides between the two
architectures (the same BATADAL attack samples), but the
attacker's degree of freedom differs. In the centralised setting,
the attacker chooses a single $\delta \in \mathbb{R}^{43}$ without
zone constraints. In the edge setting, the attacker's reach into
each zone is constrained by the zone-feature partition and the
propagation map $\iota_{ij}$. Campaign analysis
(Section~\ref{subsec:comp_validation},
Table~\ref{tab:campaign_hoe}) confirms that the BATADAL attacks
observed empirically target exactly one zone per campaign, which
matches the single-zone disturbance class used by the compositional
bound and provides empirical support for the disturbance-class
scope assumed in Eq.~\eqref{eq:comp_bound}.
 
\paragraph{Synthesis.}
On the BATADAL benchmark with per-feature RMS normalisation in
place, the three-zone edge architecture is preferable to the
centralised AE on every framework axis: $\sim 3\times$ higher
absorption capacity at $\tau = 0.60$, a $1.45$\% faster
detection time, and a localised remediation path. Both architectures share the
same recovery horizon, the same absorption-degradation divergence
under adversarial training, and the same single-zone exposure
matching the campaign structure of the benchmark. The
compositional framework provides a single integrated view of these
trade-offs that node-level or aggregate evaluation cannot deliver:
the headline edge-vs-centralised numbers are not visible from any
single per-component score, and the localisation of the
adversarial-training failure to $f_2$ is the operational content
of Eq.~\eqref{eq:binding_mismatch} applied to a real
detector deployment.

\section{Resilience-Aware IDS Design: Implications for Practice and Certification}
\label{sec:design}
 
The formal and empirical results of the preceding sections have
direct consequences for how anomaly-based IDS should be designed,
evaluated, and certified in ICS and OT environments. This section draws those consequences explicitly. The two subsections that follow address practitioners and standards bodies respectively.
 
\subsection{Implications for IDS Architecture}
\label{subsec:arch}
 
Three architectural principles follow directly from the
resilience-theoretic framework and the empirical findings
of Section~\ref{sec:empirical}.
 
\paragraph{1. Prioritise nodes on high-coupling paths,
not merely low-performing nodes.}
Standard IDS design practice targets the weakest detector,
the node with the lowest individual robustness score, for
upgrade or replacement. Eq.~\eqref{eq:binding_mismatch}
shows this is insufficient: the binding constraint on
system-level absorption capacity is determined by the ratio
$\hat{\mathcal{A}}(f_j) / W(\pi, j)$, not by
$\hat{\mathcal{A}}(f_j)$ alone. A node with moderate
absorption capacity situated on a tightly coupled path
(high $W(\pi, j)$) may impose a tighter system-level bound
than a weaker node that is poorly connected. The practical
implication is that IDS hardening budgets should be allocated
to nodes identified by the compositional bound
(Eq.~\eqref{eq:comp_bound}), not by individual node
rankings. The edge-zone deployment of~\citet{Somma2024}
provides the topological input needed for this calculation:
the PLC-motivated zone division and the hydraulically estimated
coupling weights $\hat{w}_{12} = 0.422$, $\hat{w}_{23} = 0.319$
(Eq.~\eqref{eq:coupling_values}) allow the bound to be computed
directly from the WDS network structure, without additional
modelling assumptions.
 
\paragraph{2. Calibrate detection thresholds to the
degradation function, not to a fixed operating point.}
Current practice sets a detection threshold once, either
at a fixed false-positive rate on clean data, or at the
$F_1$-maximising point on a validation set, and holds it
constant across all operating conditions. The degradation
function $\mathcal{G}_f$ (Eq.~\eqref{eq:degradation})
reveals why this is inadequate: two detectors with similar
scalar evaluation metrics may have radically different
degradation profiles under adversarial stress, as demonstrated
by the divergence between $f_{\mathrm{adv}}$ and
$f_{\mathrm{orig}}$ in Table~\ref{tab:resilience_profiles}.
Adversarial training raises the per-sample HoE distribution
yet lowers the framework's absorption capacity at the
operational threshold: $\hat{\mathcal{A}}(\tau{=}0.60)$ drops
from $0.0704$ to $0.0471$ centrally, and from $0.088$ to $0$
at the binding edge node $f_2$
(Tables~\ref{tab:resilience_profiles}, \ref{tab:comp_absorb}).
The per-sample metric moves in the opposite direction from the
composite metric; both move quietly behind a nearly unchanged
clean-data $F_1$. This is the provably fundamental
accuracy--robustness trade-off~\citep{Tsipras2019} acting
through autoencoder anomaly detectors~\citep{Beggel2019}: a
harder reconstruction objective forces a looser detection
threshold, and the looser threshold dominates the operational
outcome. A detector with a shallow degradation function should
operate at a lower threshold in high-risk periods, accepting
more false positives to retain meaningful detection capability
under moderate adversarial pressure; a detector with a steep
degradation function should trigger a fallback response as
soon as any anomalous signal appears, because performance
degrades rapidly beyond the absorption threshold. Calibrating
thresholds to $\mathcal{G}_f$ therefore requires that
$\mathcal{G}_f$ be estimated as part of the deployment
evaluation, not just individual node $F_1$ or certified radius.
 
\paragraph{3. Build in recovery mechanisms aligned with
the recovery trajectory.}
The recovery trajectory $\recov(f, t_0, \dist)$
(Eq.~\eqref{eq:recovery_traj}) and its scalar summaries
$T_r$ and $\omega_r$ expose whether a detector self-restores
after attack cessation or requires intervention. The
empirical setup of Section~\ref{subsec:recovery_empirical}
pairs a stateless AE with a post-hoc smoothing filter
yielding a structural $T_r = w - 1 = 6$\,h
(Eq.~\eqref{eq:Tr_filter}) and $\omega_r = 1$, shared across
the single-node and edge deployments. Architectural
distribution shortens attack detection latency $S_{\mathrm{TTD}}$
(Table~\ref{tab:recovery_comparison}: $0.9701$ vs.\ $0.9556$
for centralised) but does not change $T_r$, which is fixed by
the filter window. Because the detector resets on a
filter-bounded horizon and retains no longer-term memory, a
persistent slow-drift attack that stays below the detection
threshold at each time step is not surfaced at all --
$T_r$ is the wrong axis to interrogate that attack. A
resilience-aware design therefore complements the smoothed
detector with a stateful monitoring layer, a sliding-window
accumulator or sequential change-point detector, whose role
is explicitly to surface attacks with slow recovery
trajectories that smoothed detectors cannot detect. The
recovery formalism provides the diagnostic criterion: any
operational scenario in which the relevant attack signature
persists beyond the filter horizon $T_r$ requires a recovery
mechanism explicitly designed for that longer horizon.
 
\subsection{Implications for Certification and Standards}
\label{subsec:certification}
 
Current ICS security standards assess components in isolation.
IEC~62443-3-3, the system security requirements standard,
defines Security Levels (SL~1--4) as properties of individual
zones or components, with no formal mechanism for deriving a
system-level security claim from a network of components with
known coupling topology. NIST SP~800-82 (Guide to OT Security)
similarly addresses individual system components and their
configuration, not the system-level consequences of coupling
between detectors~\citep{NIST2023}. The compositional result
of this paper directly challenges the sufficiency of
component-level evaluation for ICS certification in two
respects.
 
\paragraph{Component certification does not identify the
binding constraint.}
Eq.~\eqref{eq:comp_bound} establishes that the binding
constraint on system-level resilience is the ratio
$\hat{\mathcal{A}}(f_j)/W(\pi, j)$, not $\hat{\mathcal{A}}(f_j)$
alone.
Two consequences for component-level certification follow. First,
the binding node along a path, the one whose hardening
actually tightens the system bound, need not be the
minimum-capacity node, and its identification requires the
coupling structure $\{w_{ij}\}$ and the path topology, both of
which are invisible to component-level evaluation
(Eq.~\eqref{eq:binding_mismatch}). Hardening
budgets allocated by node-level ranking alone are therefore
generically misdirected: per-node spending on a non-binding
node has no effect on $\mathcal{A}_{\mathrm{sys}}(\pi, \tau)$.
 
Second, hardening interventions assessed only at the component
level can be globally harmful even when they appear locally
beneficial on per-sample metrics. The empirical results of
Section~\ref{subsec:comp_validation} document this directly:
adversarial training of the binding node $f_2$, a standard
hardening intervention, raised the per-sample evasion
threshold (median HoE rose from $0.123$ to $0.218$ at full
features, Table~\ref{tab:comp_absorb}) yet collapsed the
system-level bound from $0.209$ to $0$ (per-feature RMS units,
Table~\ref{tab:comp_bound_result}), because the harder
reconstruction objective increased the base false-negative
rate from $7.5\,\%$ to $31.3\,\%$ and the loosened
detection threshold dominated $\hat{\mathcal{A}}(\tau)$
through the compositional bound. A hardening intervention
assessed only at per-sample HoE distributions would appear to
improve security; assessed at the operational threshold via
the framework's $\hat{\mathcal{A}}(\tau)$ and the compositional
bound, it demonstrably worsens it. This has a direct
implication for IEC~62443 compliance: the hardening priority
of detection components must be determined from the
coupling-aware compositional bound for the as-deployed network
topology, not from per-node rankings; and component-level
evaluations of hardening interventions must report
$\hat{\mathcal{A}}(\tau)$ at the operational threshold, not
only per-sample HoE distributions, to capture the
absorption-degradation divergence that propagates into the
system bound.
 
\paragraph{Degradation functions must be part of the
evaluation record.}
IEC~62443 and NIST CSF~2.0 both specify resilience
requirements in terms of recovery time objectives and
availability thresholds, but neither requires that the
degradation profile of an ML-based component under adversarial
stress be characterised as part of the component's security
documentation. The empirical result that $f_{\mathrm{adv}}$
and $f_{\mathrm{orig}}$ have nominal recall on the clean test set high above the threshold, yet markedly
different degradation functions (Table~\ref{tab:resilience_profiles})
implies that the current documentation requirement, report
accuracy and false positive rate on a clean test set, is
insufficient for ML-based anomaly detectors in safety-critical
OT environments. A standards-compatible extension would require
the degradation function $\hat{\mathcal{G}}_f(k)$ to be
reported e.g. at a minimum of two feature-availability levels as
part of the component's security evaluation record, alongside
the standard confusion matrix metrics.

\section{Conclusion}
\label{sec:conclusion}
 
This paper has argued that adversarial robustness in ICS
anomaly detection is a specific instantiation of system
resilience, and that formalising this connection yields
constructs, results, and evaluation criteria that the
existing adversarial ML literature cannot provide. We close
by restating the three contributions precisely.
 
The \textbf{first contribution} is a formal mapping of four
resilience-theoretic constructs, disturbance class,
absorption capacity, recovery trajectory, and degradation
function, onto the adversarial ML setting for OT anomaly
detection. Each mapping demonstrates that the corresponding
standard adversarial robustness metric is a degenerate
special case: the $\ell_p$-norm budget is a single-step,
semantically unconstrained disturbance class that excludes
the slow-drift and staged spoofing attacks characteristic of
real ICS adversaries; the certified radius is a point
evaluation on the absorption curve rather than the full
profile; the binary robust/not-robust classification collapses
the degradation function to a step function; and
accuracy-under-attack evaluation has no analogue of the
recovery trajectory, a structural absence that renders it
blind to temporal attack dynamics. The formal definitions
(Sections~\ref{subsec:disturbance}--\ref{subsec:degradation})
provide the vocabulary for evaluating and certifying IDS
robustness at the level of granularity that operational ICS
deployment requires.
 
The \textbf{second contribution} is a compositional resilience
bound (Eq.~\eqref{eq:comp_bound}) for heterogeneous ICS
detection networks. Modelling the network as a directed graph
of detection nodes with edge-weighted coupling strengths, the
bound establishes that the system-level absorption capacity
along any attack path is at most the minimum of each node's
absorption capacity divided by the cumulative coupling weight
to that node. This result has two operationally significant
consequences. First, the binding constraint on system-level
resilience is the ratio
$\hat{\mathcal{A}}(f_j)/W(\pi, j)$, not $\hat{\mathcal{A}}(f_j)$
alone; the binding node along a path is therefore the node
whose coupling-adjusted absorption capacity is lowest, not the
minimum-capacity node, and its identification requires the
network coupling structure and cannot be made from component
scores alone (Eq.~\eqref{eq:binding_mismatch}).
Second, the system bound is sensitive to changes in the
binding node's absorption capacity: an intervention that lowers
$\hat{\mathcal{A}}(f_{j^*})$ propagates through the bound and
can collapse system-level resilience even when the same
intervention raises per-sample evasion difficulty at the node.
The three-zone edge WDS deployment 
provides the topological substrate for this analysis: the
PLC-motivated zone division and hydraulically estimated
coupling weights ($\hat{w}_{12} = 0.422$,
$\hat{w}_{23} = 0.319$) constitute a physically grounded
instantiation of the network model.
 
The \textbf{third contribution} is an empirical demonstration
combining reinterpreted results and extended experiments from~\citet{Flatscher2024}
and~\citet{Somma2024} with new adversarial evasion and
adversarial training experiments on the three-zone edge
network conducted for this paper, showing that the
resilience metrics reveal operationally significant structure
that standard adversarial robustness benchmarks miss. Three
findings are particularly significant. First, the adversarially
trained detector $f_{\mathrm{adv}}$ raises the per-sample HoE
distribution but lowers the framework's absorption capacity
at the operational threshold: $\hat{\mathcal{A}}(\tau{=}0.60)$
drops from $0.0704$ to $0.0471$ centrally and from $0.088$ to
$0$ at the binding edge node $f_2$ -- a divergence between
per-sample geometry and composite operational behaviour
invisible to any scalar benchmark but resolved by the
absorption-curve and degradation-function framework, and
consistent with the provably fundamental accuracy--robustness
trade-off of adversarial ML~\citep{Tsipras2019}. Second, the
BATADAL campaign analysis (new to this paper) reveals that
each attack campaign targets sensors in exactly one zone,
independently validating the PLC-zone partition
of~\citet{Somma2024} and establishing that under a
sustained-attack threat model the operationally relevant
absorption capacity statistic is the campaign-maximum HoE
(Eq.~\eqref{eq:campaign_max}): an adversary conducting a
sustained campaign must conceal every sample, not only the
typical one. Third, adversarial training of the binding node
$f_2$, intended to raise its per-sample evasion threshold, collapsed the system-level bound from $0.209$ to $0$
(per-feature RMS units, Table~\ref{tab:comp_bound_result}),
because the harder reconstruction objective raised the base
false-negative rate from $7.5\,\%$ to $31.3\,\%$ and the
loosened detection threshold dominated
$\hat{\mathcal{A}}(\tau)$ through the compositional bound.
This is the first formal demonstration, at the system level
under a compositional threat model, that the
accuracy--robustness trade-off documented for individual
classifiers propagates into system-level fragility in
heterogeneous ICS detection networks. Applied together on the BATADAL benchmark, these metrics yield a concrete architectural verdict: the three-zone edge deployment is more resilient than the centralised detector on every framework axis, with $\sim 3\times$ higher absorption capacity at the operational threshold.
 
Taken together, these three contributions establish a
compositional resilience framework for ICS anomaly detection
in which system-level adversarial robustness is a derived
quantity, traceable to the resilience profiles of constituent
detection nodes and to the coupling structure of the deployed
network. The framework yields guarantees that are directly
interpretable in the language of the certification standards
that govern ICS deployment, and we have shown on a realistic
benchmark that it surfaces operationally significant structure
that standard adversarial evaluation does not.

Extension of the framework to broader threat models and deployment topologies is left for future work.

\appendix
\section{Derivation of the compositional bound}
\label{app:proof}
 
This appendix gives a complete derivation of the compositional
bound of Eq.~\eqref{eq:comp_bound} under the following standing
conditions, which hold for all anomaly detectors and coupling maps
considered in this paper.
 
\paragraph{Assumption: Detector regularity.}
The performance metric $\Phi(f, \cdot) : \mathcal{X} \to [0,1]$
is continuous and non-increasing in disturbance magnitude:
if $\norm{\boldsymbol{\delta}}_p \leq \norm{\boldsymbol{\delta}'}_p$
then $\Phi(f, x + \boldsymbol{\delta}) \geq
\Phi(f, x + \boldsymbol{\delta}')$ for all $x \in \mathcal{X}$.
 
\paragraph{Assumption: Existence of worst-case disturbance.}
For each node $f_{i_j}$ and budget $b \geq 0$, the set
$\dist_{i_j}(b)$ is compact and the infimum in
Eq.~\eqref{eq:absorption} is attained: there exists
$\boldsymbol{\delta}^*_{i_j}(b) \in \dist_{i_j}(b)$ such that
$\Phi(f_{i_j},\, x_{i_j} + \boldsymbol{\delta}^*_{i_j}(b)) =
\inf_{\boldsymbol{\delta} \in \dist_{i_j}(b)}
\Phi(f_{i_j},\, x_{i_j} + \boldsymbol{\delta})$.
This follows from the continuity of $\Phi$ and compactness
of $\dist_{i_j}(b)$ by the extreme value theorem.
 
\paragraph{Assumption: Coupling map regularity.}
For each directed edge $(f_{i_l}, f_{i_{l+1}}) \in E$,
the signal-propagation map
$\iota_{i_l i_{l+1}} : \mathcal{X}_{i_l} \times \R^{d_{i_l}}
\to \R^{d_{i_{l+1}}}$ satisfies
$\norm{\iota_{i_l i_{l+1}}(x_{i_l},\, \boldsymbol{\delta})}_p
\leq w_{i_l i_{l+1}} \cdot \norm{\boldsymbol{\delta}}_p$
for all $x_{i_l} \in \mathcal{X}_{i_l}$ and all
$\boldsymbol{\delta} \in \R^{d_{i_l}}$, where $w_{i_l i_{l+1}}$
is the coupling weight from Section~\ref{subsec:network_model}.
This is the contractivity condition that ensures
$w_{ij} \leq 1$ and follows from the definition of $w_{ij}$
as a Lipschitz constant of $\iota_{ij}$
(Eq.~\eqref{eq:coupling_weight}).
 
\paragraph{Derivation of the compositional bound.}
 
Let $\pi = (f_{i_1}, \ldots, f_{i_k})$ be a directed attack
path in $G$ and let $b > 0$ be an adversarial budget at the
entry node $f_{i_1}$. We proceed in four steps.
 
\medskip\noindent
\textbf{Step 1: Budget propagation along a single edge.}
 
Consider a single directed edge $(f_{i_l}, f_{i_{l+1}}) \in E$.
Suppose the attacker applies a disturbance
$\boldsymbol{\delta}_{i_l} \in \dist_{i_l}(b_l)$ at node
$f_{i_l}$, where $\|\boldsymbol{\delta}_{i_l}\|_p \leq b_l$. The induced disturbance on $f_{i_{l+1}}$ via the
coupling map is $\boldsymbol{\delta}_{i_{l+1}} =
\iota_{i_l i_{l+1}}(x_{i_l}, \boldsymbol{\delta}_{i_l})$.
By the coupling-map contractivity condition:
\begin{equation}
    \norm{\boldsymbol{\delta}_{i_{l+1}}}_p
    \;=\;
    \norm{\iota_{i_l i_{l+1}}(x_{i_l},\,
    \boldsymbol{\delta}_{i_l})}_p
    \;\leq\;
    w_{i_l i_{l+1}} \cdot
    \norm{\boldsymbol{\delta}_{i_l}}_p
    \;\leq\;
    w_{i_l i_{l+1}} \cdot b_l.
    \label{eq:prop_single}
\end{equation}
Therefore $\boldsymbol{\delta}_{i_{l+1}} \in
\dist_{i_{l+1}}(w_{i_l i_{l+1}} \cdot b_l)$: the effective
budget at $f_{i_{l+1}}$ is at most
$w_{i_l i_{l+1}} \cdot b_l$.
 
\medskip\noindent
\textbf{Step 2: Cumulative budget propagation along the path.}
 
Define the effective budget at node $f_{i_j}$ as $b_j
\triangleq \norm{\boldsymbol{\delta}^*_{i_j}}_p$, where
$\boldsymbol{\delta}^*_{i_j}$ is the induced disturbance
arriving at $f_{i_j}$ after propagation from $f_{i_1}$.
We claim $b_j \leq b \cdot W(\pi, j)$ for all
$j = 1, \ldots, k$, where $W(\pi, j)$ is the cumulative
coupling defined in Eq.~\eqref{eq:cumulative_coupling}.
 
\emph{Base case} ($j=1$): $W(\pi,1) = 1$ (empty product),
so $b_1 = b \leq b \cdot 1$. The claim holds.
 
\emph{Inductive step}: Assume $b_j \leq b \cdot W(\pi,j)$.
Applying~\eqref{eq:prop_single} to edge
$(f_{i_j}, f_{i_{j+1}})$:
\begin{equation}
    b_{j+1}
    \;\leq\; w_{i_j i_{j+1}} \cdot b_j
    \;\leq\; w_{i_j i_{j+1}} \cdot b \cdot W(\pi,\, j)
    \;=\; b \cdot W(\pi,\, j+1).
    \label{eq:inductive_step}
\end{equation}
By induction, $b_j \leq b \cdot W(\pi, j)$ for all
$j = 1, \ldots, k$.
 
\medskip\noindent
\textbf{Step 3: Node safety condition.}
 
Node $f_{i_j}$ maintains detection performance above
threshold $\tau$ under effective budget $b_j$ if and only if
$b_j \leq \mathcal{A}(f_{i_j}, \dist, \tau)$.
 
$(\Rightarrow)$: By definition of absorption capacity
(Eq.~\eqref{eq:absorption}), if $b_j \leq
\mathcal{A}(f_{i_j}, \dist, \tau)$ then
$\inf_{\boldsymbol{\delta} \in \dist_{i_j}(b_j)}
\Phi(f_{i_j}, x_{i_j} + \boldsymbol{\delta}) \geq \tau$.
In particular, $\Phi(f_{i_j}, x_{i_j} +
\boldsymbol{\delta}^*_{i_j}) \geq \tau$.
 
$(\Leftarrow)$: If $b_j > \mathcal{A}(f_{i_j}, \dist, \tau)$,
then by the non-increasing property of $\Phi$
(by the detector regularity assumption), the definition of
the supremum in Eq.~\eqref{eq:absorption},
and the existence-of-worst-case-disturbance assumption, there exists
$\boldsymbol{\delta}^* \in \dist_{i_j}(b_j)$ such
that $\Phi(f_{i_j}, x_{i_j} + \boldsymbol{\delta}^*)
< \tau$.
 
Combining with the budget bound from Step 2:
if $b \leq \mathcal{A}(f_{i_j}, \dist, \tau) / W(\pi, j)$,
then $b_j \leq b \cdot W(\pi,j) \leq
\mathcal{A}(f_{i_j}, \dist, \tau)$, so node $f_{i_j}$
is safe.
 
\medskip\noindent
\textbf{Step 4: System-level absorption capacity.}
 
The system maintains performance above $\tau$ along path $\pi$
if and only if every node $f_{i_j}$ on the path is safe.
By Step 3, this requires $b_j \leq \mathcal{A}(f_{i_j},
\dist, \tau)$ for all $j = 1, \ldots, k$. Combined with
Step 2, a sufficient condition for all nodes to be safe is:
\begin{equation}
    b \;\leq\; \frac{\mathcal{A}(f_{i_j},\, \dist,\, \tau)}
    {W(\pi,\, j)}
    \quad \text{for all } j = 1, \ldots, k,
    \label{eq:suff_cond}
\end{equation}
which holds if and only if:
\begin{equation}
    b \;\leq\; \min_{j=1,\ldots,k}\;
    \frac{\mathcal{A}(f_{i_j},\, \dist,\, \tau)}{W(\pi,\, j)}.
    \label{eq:min_cond}
\end{equation}
Taking the supremum over all $b$ satisfying the system-level
condition $\varrho_{\mathrm{sys}}(\pi, b) \geq \tau$:
\begin{equation}
    \mathcal{A}_{\mathrm{sys}}(\pi,\, \tau)
    \;=\; \sup\bigl\{b \geq 0 :
    \varrho_{\mathrm{sys}}(\pi, b) \geq \tau\bigr\}
    \;\leq\;
    \min_{j=1,\ldots,k}\;
    \frac{\mathcal{A}(f_{i_j},\, \dist,\, \tau)}{W(\pi,\, j)},
    \label{eq:bound_final}
\end{equation}
which is the compositional bound of Eq.~\eqref{eq:comp_bound}.
 
\paragraph{Why the bound is an inequality, not equality.}
The inequality in~\eqref{eq:bound_final} arises because
Eq.~\eqref{eq:suff_cond} is a \emph{sufficient} condition
for system safety, not a necessary one. Specifically, the
bound from Step 2 ($b_j \leq b \cdot W(\pi,j)$) uses the
worst-case coupling, i.e. the supremum in
Eq.~\eqref{eq:coupling_weight}, to upper-bound the
effective budget. The true propagated budget depends on the
specific disturbance $\boldsymbol{\delta}_{i_l}$ applied,
which may not saturate the coupling weight. As a result,
$\mathcal{A}_{\mathrm{sys}}$ may strictly exceed the bound.
 
\paragraph{Conditions for tightness.}
The bound is tight (equality holds in~\eqref{eq:bound_final})
under three conditions simultaneously:
\begin{enumerate}[label=(\roman*)]
    \item the coupling maps $\iota_{ij}$ are linear and
          isometric with respect to the coupling weight,
          i.e., $\norm{\iota_{ij}(x, \boldsymbol{\delta})}_p
          = w_{ij} \cdot \norm{\boldsymbol{\delta}}_p$ for
          all $x$ and all $\boldsymbol{\delta}$;
    \item the attacker can independently choose the
          disturbance at each node so as to simultaneously
          saturate each coupling link; and
    \item the minimum in~\eqref{eq:min_cond} is attained at
          a unique node $f_{i_{j^*}}$ (so that a small
          increase in $b$ beyond the bound necessarily
          compromises $f_{i_{j^*}}$).
\end{enumerate}
When the coupling maps are strictly contractive
($\norm{\iota_{ij}(\cdot, \boldsymbol{\delta})}_p <
w_{ij} \cdot \norm{\boldsymbol{\delta}}_p$ for some
$\boldsymbol{\delta}$), the true $\mathcal{A}_{\mathrm{sys}}$
will strictly exceed the bound, and the bound is conservative.
Characterising the gap between the bound and the true value
as a function of the coupling topology is an open problem.

\section*{CRediT authorship contribution statement}
\textbf{Branka Stojanović:} Conceptualization of this study, Methodology, Formal analysis, Investigation, Writing -- Original Draft.
\textbf{Andreas Flatscher:} Investigation, Writing -- Review \& Editing.
\textbf{Michael Somma:} Validation, Writing -- Review \& Editing.

\end{document}